\documentclass[a4paper,11pt,headings=big,DIV=12]{scrartcl}
\usepackage{amsmath,amssymb}
\usepackage[colorlinks,linkcolor=mygray,urlcolor=blue,citecolor=blue]{hyperref}
\usepackage{tcolorbox}
\definecolor{mygray}{gray}{0.2}
\definecolor{mypink1}{rgb}{0.9, 0.2, 0.6}
\usepackage{graphicx}
\addtokomafont{disposition}{\rmfamily\boldmath}
\usepackage[utf8]{inputenc}
\usepackage{enumerate} 
\usepackage{slashed}
\usepackage{cite} 
\usepackage{comment}
\usepackage{multirow}
 \usepackage{bbold}
\allowdisplaybreaks
\newcommand{\vev}[1]{\langle #1 \rangle} 
\newcommand{\state}[1]{|#1\rangle}
\newcommand{\astate}[1]{\langle #1|}
\newcommand{\matel}[3]{\langle #1|#2|#3\rangle}
\newcommand{\al}{\alpha}
\newcommand{\be}{\beta}
\newcommand{\ga}{\gamma}
\newcommand{\de}{\delta}
\newcommand{\la}{\lambda}
\newcommand{\eps}{\epsilon}

\newcommand{\Rtr}[2]{\Theta}

\newcommand*{\mathcolor}{}
\def\mathcolor#1#{\mathcoloraux{#1}}
\newcommand*{\mathcoloraux}[3]{%
  \protect\leavevmode
  \begingroup
    \color#1{#2}#3%
  \endgroup
}

\begin{document}

\begin{flushright}
\begin{tabular}{l}
CP3-Origins-2016-042 DNRF90
\end{tabular}
\end{flushright}
\vskip1.5cm

\begin{center}
{\Large\bfseries \boldmath A brief Introduction to Dispersion Relations and Analyticity}\footnote{Based on a three-hours blackboard lecture  
given at the school in Dubna, Russia 18-20 July 2016 "Strong fields and Heavy Quarks".  
Lectures to appear in upcoming proceedings of the school.
Added discussion of anomalous thresholds and second type singularities.}
\\[0.8 cm]
{\Large%
 Roman Zwicky
\\[0.5 cm]
\small
 Higgs Centre for Theoretical Physics, School of Physics and Astronomy,\\
University of Edinburgh, Edinburgh EH9 3JZ, Scotland 
} \\[0.5 cm]
\small
E-Mail:
\texttt{\href{mailto:roman.zwicky@ed.ac.uk}{roman.zwicky@ed.ac.uk}}.
\end{center}

\bigskip
\pagestyle{empty}

\begin{abstract}\noindent
In these lectures we provide a basic introduction into the topic of dispersion relation and analyticity. 
 The properties of 2-point functions are discussed in some detail from the viewpoint of the 
 K\"all\'en-Lehmann and general 
dispersion relations.  The Weinberg sum rules figure as an application.  The analytic structure 
of higher point functions in perturbation theory are analysed through the Landau equations  and the 
Cutkosky rules. 
 \end{abstract}

\newpage

\setcounter{tocdepth}{3}
\setcounter{page}{1}
\tableofcontents
\pagestyle{plain}

\section{Prologue}

Dispersion relations are a powerful non-perturbative tool which have originated in classical electrodynamics 
in the theory of Kramers-Kronig dispersion relations. Analytic properties  follow from causality and the use
of Cauchy's theorem allows to obtain the real part of an amplitude 
from the knowledge of the 
imaginary part which is often better accessible. 
This is the idea of the S-matrix program from the fifties and sixties. 
Dispersion relations are sparsely discussed in modern textbooks as the focus is on other aspects of 
Quantum Field Theory (QFT).  
There are some excellent older textbooks on analyticity  e.g. \cite{S-matrix,Todorov,Barton}, some 
modern textbooks devote some chapters to the topic e.g.\cite{IZ,Weinberg1}, 
as well as some lecture notes   \cite{deRafael}.
I would hope that a student who has followed an introductory course on QFT or has read some  chapters
of a QFT textbook would be able to largely follow the presentation below.

\subsection{Introduction}
\label{sec:prologue}

 In the fifties and sixties QFT 
has found a big success in describing quantum electrodynamics (QED) thanks to the successful 
renormalisation program carried out by Dyson, Feynman, Schwinger, Tomonaga and others \cite{Schweber}.
The description of the strong force with QFT proved to be difficult and there was some prejudice that 
a solution outside field theory had to be found. Two such approaches are 
dispersion theory using analytic properties \cite{S-matrix} (Heisenberg, Chew, \dots) and  Wilson's operator product expansion \cite{Wilson69}.
As Weinberg remarks in his book \cite{Weinberg1} both of these approaches later became a part of QFT! 
By analytic properties we mean analyticity in the external momenta.
In QFT analytic continuation is inherent in the field description (second quantisation). 
Let us remind ourselves how this is related to scattering matrix elements of particles. 
\begin{figure}
\begin{center}
\includegraphics[width=2.5in]{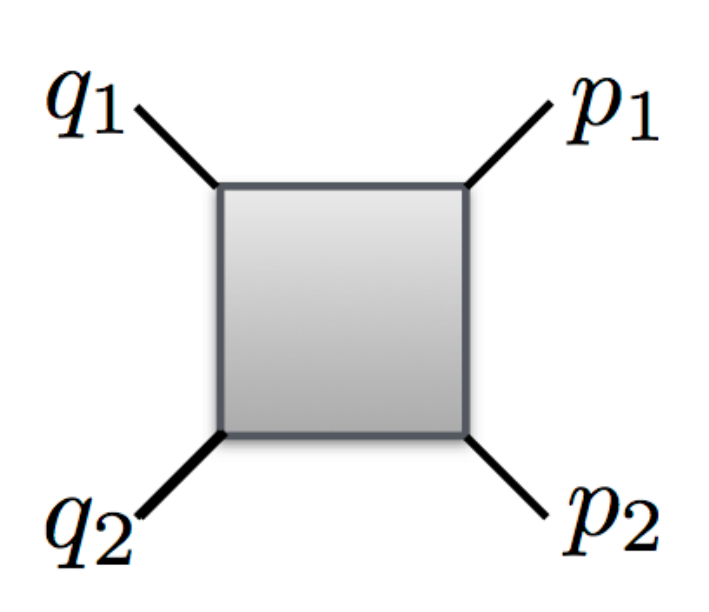}
\end{center}
\caption{\small Schematic diagram for $2\to 2$ scattering corresponding to the matrix element in Eq.~\eqref{eq:Smatrix}. Time goes from left to right.}
\label{fig:2to2}
\end{figure}

A primary goal of particle physics is to describe scattering of $n$-particles via the so-called S-matrix.
For the scattering of $ 2 \to 2$ particles this reads (cf. Fig.\ref{fig:2to2})
\begin{equation}
\label{eq:Smatrix}
\phantom{a}_{\rm out}\astate{p_1,p_2} q_1 q_2      \rangle_{\rm in}  = \phantom{a}_{\rm out}\astate{p_1,p_2} 
S |q_1 q_2      \rangle_{\rm out}  \;,
\end{equation}
where we have assumed the particles to be of spin $0$. In the case where they are all of equal mass this implies the following on-shell conditions: $p_1^2 = p_2^2 = q_1^2 = q_2^2 = m^2$. 
Hence one might wonder how  analytic properties come into play. 
The answer is through the celebrated \emph{Lehman-Symmanzik-Zimmermann (LSZ) formula} whose 
 derivation can be found in most textbooks e.g. \cite{IZ}.
For our case it reads
\begin{eqnarray}
\label{eq:LSZ}
\phantom{a}_{\rm out}\astate{p_1,p_2} q_1 q_2      \rangle_{\rm in}  &=& - (i Z^{-1/2})^4 \int_{x_1,x_2,y_1,y_2} 
e^{- i ( q_1 \cdot x_1 + q_2 \cdot x_2- p_1\cdot  y_1  - p_2\cdot  y_2   )} K_{x_1} K_{x_2} K_{y_1} K_{y_2}  \times\nonumber  \\ 
& & \vev{T \phi(x_1) \phi(x_2)\phi^\dagger(y_1)\phi^\dagger(y_1)} + \text{disconnected terms} \;,
\end{eqnarray}
where $\int_x = \int d^4 x$ hereafter, $T$ is the time ordering, $\vev{\dots}$ is the vacuum expectation value (VEV), 
 the quanta are assumed to carry a charge (complex conjugation for outgoing particle), $K$ is the Klein-Gordon operator 
$K_{x_1} = \Box + m^2 \to -(q_1^2-m^2)$ and the $Z$ factor results from the \emph{asymptotic 
condition}, 
\begin{equation}
\label{eq:asym}
\matel{0 }{\phi(x)}{q_1}  \stackrel{x_0 \to \mp \infty}{\to}   Z^{1/2} \matel{0 }{\phi_{\text{in(out)}}(x)} {q_1} \;,
\end{equation}

The asymptotic condition is the key idea of the LSZ-approach. Namely that when the particles are well separated 
from each other all that remains is the self-interaction which is parameterised by the renormalisation factor $Z$. The field $\phi$ is what is known as an interacting field whereas $ \phi_{\text{in(out)}}$ are free fields in which 
case the right-hand side of the equation above equals $\sqrt{Z /(2\pi)^3}e^{-i q_1 \cdot x}$.\footnote{\label{foot:inter} The LSZ formalism, in its elegancy and efficiency, also allows for the description of composite particles. For example  a pion of $SU(2)$-isospin quantum number $a$ may be described by 
$\phi \to \varphi^a =  \phi \bar q T^a \ga_5 q$ in the sense that $\matel{0}{   \varphi^a }{\pi^b} = g_\pi \de^{ab}$. 
In such a case $\varphi^a$ is referred to as an interpolating field.}\footnote{It is crucial that 
this condition is only imposed on the matrix element (weak topology) as otherwise one runs into 
Haag's theorem \cite{Haag} which states that any field which is unitarity equivalent to a free field 
is itself a free field.}
 The disconnected part corresponds, for example, to the case where particle $q_1 \to p_1$ 
and $q_2 \to p_2$ without any interaction 
which is of no interest to us. 
From \eqref{eq:LSZ} we conclude  that 
\begin{itemize}
\item[a)] The scattering of $n$-particles  ($n = n_{\rm in } + n_{\rm out}$) is described by $n$-point functions (or $n$-point correlators).  The 
study of the latter is therefore of primary importance. 
\item[b)] The $n$-point correlators are functions of the external momenta e.g. $p_{1,2}^2, q_{1,2}^2, 
p_1 \cdot p_2 ,\dots $. First and foremost they are defined for real values or more precisely for real values 
with a small imaginary part e.g. 
$p_1^2 = {\rm Re}[p_1^2]  + i0$.\footnote{In perturbation theory (PT) 
the reality of the momenta is implicitly used when shifting momenta (e.g. completing squares for example).}
From there they can be analytically continued into the complex plane. 
Hence it is the second quantisation, describing particles with fields, that allows us to 
 go off-shell for correlation functions.\footnote{In it's most standard formulation string theory is first quantised and does not allow this analytic continuation. String field 
theory does exist but is less developed than first quantised string theory for technical reasons.}
\end{itemize}

The course consists of three parts. Analytic properties of 2-point functions (section \ref{sec:2pt}),
which comes with definite answer in terms  
of the non-perturbative K\"all\'en-Lehmann spectral representation. Applications of 2-point function 
in section \ref{sec:app}. Last  a short discussion of the analytic properties of 
higher point function in perturbation theory (PT) e.g. Landau equations and Cutkosky rules in section \ref{sec:higher}. 

\section{2-point Function}
\label{sec:2pt}

\subsection{Dispersion Relation from $1^{\text{st}}$-principles: K\"all\'en-Lehmann Representation}
\label{sec:KL}

Let us define the Fourier transform of the 2-point correlator as follows 
\begin{equation}
\Gamma(p^2) = i \int_x e^{i p \cdot x} \vev{T \phi(x) \phi^\dagger(0)} \;.
\end{equation}
What determines the analytic structure of $\Gamma(p^2)$?  By analytic structure we mean the singularities e.g. poles, branch points and the associated branch cuts. The K\"all\'en-Lehmann  
representation \cite{K,L} gives a very definite answer to this question. The presentation is straightforward 
and can be found in most textbooks e.g. \cite{Weinberg1}.

\begin{figure}
\begin{center}
\includegraphics[width=4.6in]{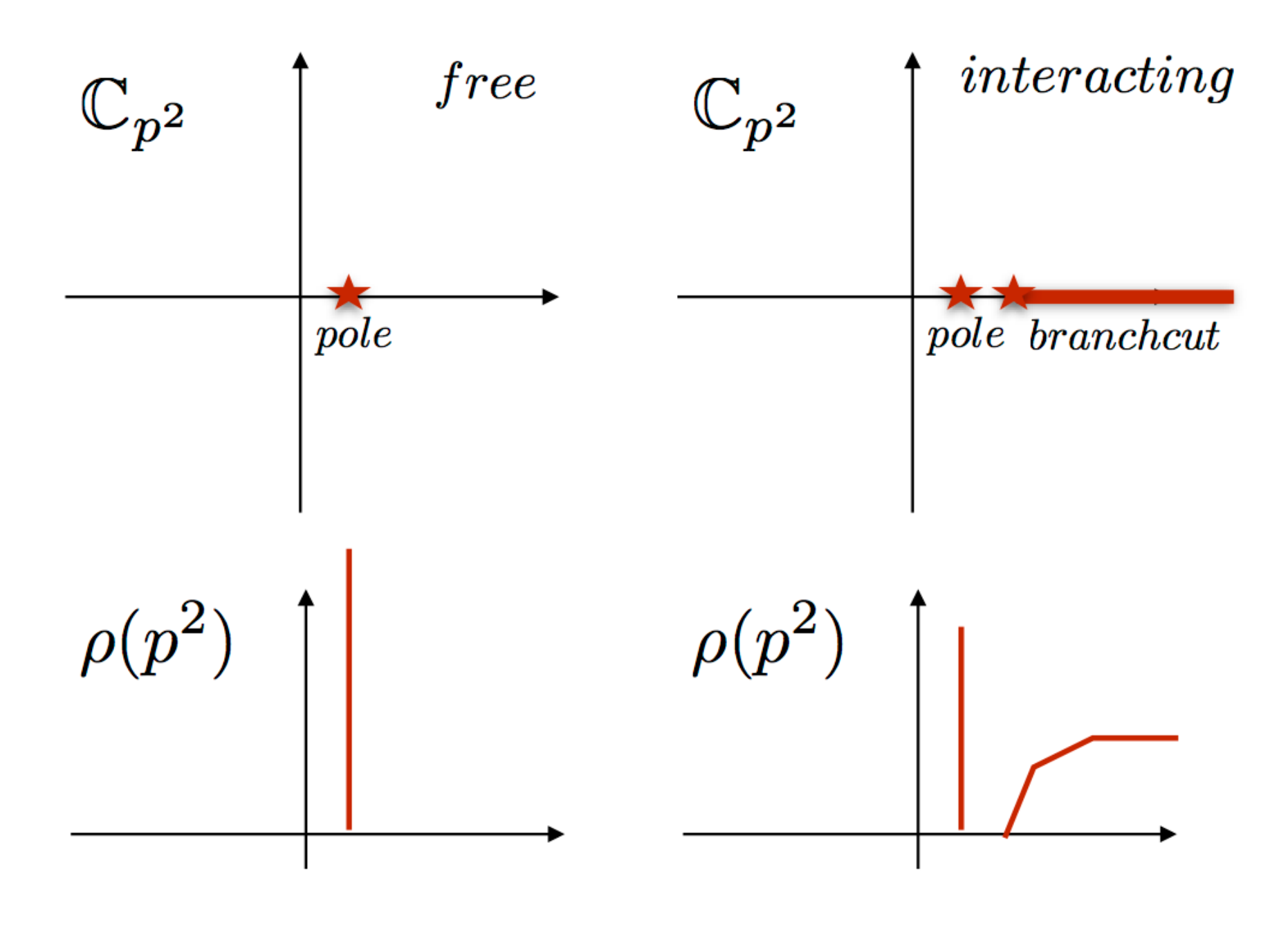} \; 
\end{center}
\caption{\small (left) analytic structure for free field theory propagator with spectral function underneath 
(right) idem for an interacting theory 
with a stable 1-particle state and a multiparticle-threshold}
\label{fig:analytic-structure}
\end{figure}

The 2-point function in the free and interacting case can be written as
\begin{equation}
\Gamma(p^2) = \left\{
\begin{array}{ll}
\label{eq:Ga}
\frac{1}{m^2 - p^2 - i 0} = - \Delta_F(p^2,m^2) & \text{free} \\[0.3cm]
\frac{ Z(\la)}{m^2 - p^2 - i 0} + f(\la,p^2) & \text{interacting}
\end{array}
\right.  \;.
\end{equation}
The function $Z(\la)$ and $f(\la,p^2)$, where $\la$ is  the coupling constant 
e.g. ${\cal L}_{\rm int} = \la \phi^3 + {\rm h.c.}$, obey  
\begin{equation}
Z(\la) \stackrel{\la \to 0}{\to} 1 \;, \quad f(\la,p^2)\stackrel{\la \to 0}{\to} 0 \;,
\end{equation}
in order to reproduce the  free field theory limit.
In what follows it is our goal to determine the properties of $f(\la,p^2)$ in more detail. 
At the end of this section we are going to make remarks about the possible ranges of the 
$Z(\la)$-function. The first lesson to be learnt from the free field theory case is that it  is 
the mass (i.e. the spectrum) which determines the analytic properties cf. Fig.~\ref{fig:analytic-structure}(left). As we shall see this 
generalises to the interacting case.

For technical reason it is advantageous to first study the positive frequency distribution 
\begin{equation}
\label{eq:T}
\vev{ \phi(x) \phi^\dagger(0)} = \left\{
\begin{array}{ll}
 \Delta_+(x^2,m^2) = \int \frac{d^4 p}{(2 \pi)^3} e^{- i p \cdot x}\delta^+(p^2-m^2)  & \text{free} \\[0.3cm]
(*) & \text{interacting}
\end{array}
\right.  
\end{equation}
where $\delta^+(p^2-m^2) \equiv \delta(p^2-m^2)\theta(p_0)$  assures that energies are positive and 
that the momenta are on the mass-shell. It is the quantity $(*)$ that we intend to study. First we  
use the formal decomposition of the identity into a complete set of states $\mathbb{1} = 
\sum_n \state{n}\astate{n}$ which follow from unitarity. Inserting this relation and using 
translation invariance one gets
\begin{equation}
(*) = \sum_n e^{-i p_n \cdot x} |\underbrace{ \matel{0}{ \phi(0)}{n(p_n)}}_{ \equiv f_n}|^2 \;. 
\end{equation}
Further using $1 = \frac{1}{(2 \pi)^4}\int_p e^{- i p \cdot x}  \int_x e^{ i p \cdot x}$ and interchanging 
the $\sum_n$ and the $\int_x$\footnote{We will come back to these interchanges which are ill-defined 
when there are UV-divergences.} leads to
\begin{equation}
\label{eq:spec}
(*) = \int d^4 p e^{- i p \cdot x} \underbrace{ \equiv \sum_n \de(p-p_n) |f_n|^2}_{(2 \pi)^{-3} \rho(p^2) \theta(p_0) }  \;,
\end{equation}
where $\rho(p^2)$ is known as the \emph{spectral function}, $(2\pi)^{-3}$ a convenient normalisation factor 
and $\theta(p_0)$ assures positive energies which come from the positive energy condition on the external momentum.
 Upon using $\int_p F(p) =  \int_p \int ds \de(s -p^2) F(s)$ and exchanging the $ds$ and $d^4 p$ integration 
 one finally gets
 \begin{equation}
 (*)  = \int_0^\infty ds \rho(s) \Delta_+(x^2,s) \;,
 \end{equation}
 a spectral representation.
 
From \eqref{eq:Ga} and \eqref{eq:T} it seems plausible that  
this spectral representation generalises to the time ordered 2-point function as follows
\begin{tcolorbox}
\begin{equation}
\label{eq:KL}
 \Gamma(p^2) = \int_0^\infty ds \rho(s) (-\Delta_F(s,p^2)) = \int_0^\infty ds \frac{\rho(s)}{s-p^2 - i0} \;.
\end{equation} 
Eq.~\eqref{eq:KL} is  referred to as the 
\emph{K\"all\'en-Lehmann (spectral) representation}.
\end{tcolorbox}
 At this stage we can make many relevant comments.
\begin{enumerate}
\item The K\"all\'en-Lehmann  representation is a special case of dispersion relation. 
It shows that dispersion representation follow from first principles in QFT.
\item The analytic properties of $\Gamma(p^2)$ are in one-to-one correspondence with 
the spectrum of the theory which is the answer to the question what determines the 
analytic properties of the 2-point function. 
 Hence for the 2-point function there are no other singularities on the first sheet (known as 
the physical sheet)\footnote{More precisely the 2-point function is at first defined for real $p^2 + i0$ with 
$p^2 \in \mathbb{R}$. Analytic continuation which is unique from an interval proceeds through the upper half-plane to the left and passes below zero 
for real $p^2$ below the singlarities on the positive real line.}
  other than on the positive real axis determined by the spectrum. The analytic structure is depicted in Fig.~\ref{fig:analytic-structure}(right). 
  An example of an unphysical singularity (not on the physical sheet) is given 
  in section \ref{sec:example}.
  \item The spectral function $\rho(s) \geq 0 $ is positive definite as a direct consequence of unitarity.  
[As a homework question you could try to show that for a non-unitary theory with negative normed states 
(i.e. $\langle gh|gh \rangle = -1$ where ``gh" stands for ghost)
$\rho(s)$  loses positive definiteness.] 
\item Often the spectral function is decomposed into  a pole part\footnote{When the particle becomes unstable 
and acquires a width then the pole wanders on the second sheet since the principle that there are no singularities on the physical sheet holds up e.g. \cite{S-matrix}. This would have been an interesting additional topic which 
we can unfortunately not cover in these short lectures.} 
\begin{equation}
\label{eq:rho}
\rho(s) = Z \delta(s-m^2)  + \theta(s-s_0) \sigma(s)
\end{equation}
and continuum part $\sigma(s)$.  The latter is the concrete realisation of the  function $f(\la,s)$ in \eqref{eq:Ga}.
In many applications $f_0$, the residue of the lowest state, 
\begin{equation}
\label{eq:NP}
\Gamma(p^2) = \frac{|f_0|^2}{m^2 -p^2 -i0}  + \int_{s_0}^\infty \frac{\sigma(s)}{m^2 -p^2 -i0} \;,
\end{equation}
is the non-perturbative quantity that is  to be extracted. The left-hand side is computed and the 
$\sigma$-part is then either estimated or suppressed by applying an operation  to the equation. 
This technique is the basis of QCD sum rules \cite{SVZ} and lattice QCD \cite{DeGrand:2006zz} extraction 
of low-lying hadronic parameters. 
In the former case the $\sigma$-part is suppressed by a Borel-transformation and in lattice QCD $\sigma$-part is exponentially suppressed in euclidian time. 
\item The K\"all\'en-Lehmann representation straightforwardly applies to the case of a non-diagonal 
correlation function e.g. $\vev{\phi_A(x) \phi_B^\dagger(0)}$ but clearly  positive definiteness is, in general, 
lost since $|f_n|^2 \to f_n^A (f_n^B)^*$.
\item As promised we return to the issue of interchanging various sums and integrals. This is of no 
consequence as long as there are no UV-divergences. As is well-known most field theories show 
UV-divergences so care has to be taken. UV-divergences demand regularisations and a prescription 
to renormalise the ambiguities which arise from removing the infinities.   
There are two ways to formally handle this problem. First, assuming a logarithmic divergence, we may
amend  \eqref{eq:KL} as 
\begin{equation}
\label{eq:subtract}
\Gamma(p^2) = \int_0^{\Lambda_{\rm UV}} ds \frac{\rho(s)}{s-p^2 - i0} + A  \;,
\end{equation}
where the so-called subtraction constant is adjusted to cancel the logarithmic divergence 
coming form the integral: $A = A_0  \ln (\Lambda_{\rm UV}^2/\mu_0^2) + A_1$ with $\mu_0$ being 
some arbitrary reference scale. The constant  $A_1$ has either to be taken from experiment in the case where $\Gamma(p^2) $ is physical (which implies scheme-independence) or is dependent on the scheme. The dependence in the latter case has to disappear 
when physical information is extracted from $\Gamma(p^2)$. 
A more elegant way, in my opinion, 
 is to handle the problem with
a once subtracted dispersion 
\begin{equation}
\label{eq:subtractDR}
\Gamma(p^2) = \Gamma(p_0^2) + (p^2-p_0^2)
\int_0^{\infty} ds \frac{\rho(s)}{(s-p^2 - i0)(s-p_0^2)}   \;.
\end{equation}
It is observed that the integral is now convergent due to the extra $1/(s-p_0^2)$ factor. 
The same remarks apply to $\Gamma(p_0^2)$ as for the previously discussed $A_1$. 
To derive the above expression one writes an unsubtracted dispersion relation for 
$\Gamma(p^2)$ and $\Gamma(p_0^2)$ separately takes the difference and 
combines the fraction. They key point is that the divergent parts are the same and cancel each other.
\item Following the presentation in Weinberg's book \cite{Weinberg1}:
 imposing the canonical commutation relation $[ \partial_t \phi^\dagger(x) , \phi(0)]_{x_0=0} = 
-i \de( \vec{x})$  (in $\hbar = 1$ units) leads to the sum rule
\begin{equation}
\label{eq:SR}
\int_0^\infty ds \rho(s) = 1 \;,
\end{equation}
from where one deduces that:
\begin{itemize}
\item[-] $Z= 1$  for a free theory
\item[-] $0 \leq Z \leq 1$  for an interacting theory 
\item[-] $Z = 0$ if $\phi$ is a confined field 
\end{itemize} 
The last case does not follow directly from \eqref{eq:SR} but is an important result due to Weinberg. 
An example  is given by the quark propagator for which we do not expect a residue since
it is a confined (coloured) particle. The fact that $Z_{\rm quark} \neq 0$ in each order 
in PT is  a sign that the latter is not suited to describe the phenomenon of confinement.
\item By using causality, i.e. $\vev{ [\phi(x), \phi^\dagger(0) ]}=0$ for $x^2 < 0$ spacelike, it follows that 
$\bar{\rho}(s) = \rho(s)$ where $\bar{\rho}(s)$ is the antiparticle spectral function associated with 
$\Delta_-(x^2,m^2) = \vev{\phi^\dagger(x) \phi(0)}$. This is a special case of the CPT theorem. 
Related to this matter it was Gell-Mann, Goldberger and Thirring \cite{GGT} in 1954 who derived analyticity properties
from causality, for $\ga + N \to \ga +N$, justifying dispersion relations from a non-perturbative viewpoint.
\end{enumerate}

\subsection{Dispersion Relations and Cauchy's Theorem}
\label{sec:cauchy}

\begin{figure}
\begin{center}
\includegraphics[width=4.0in]{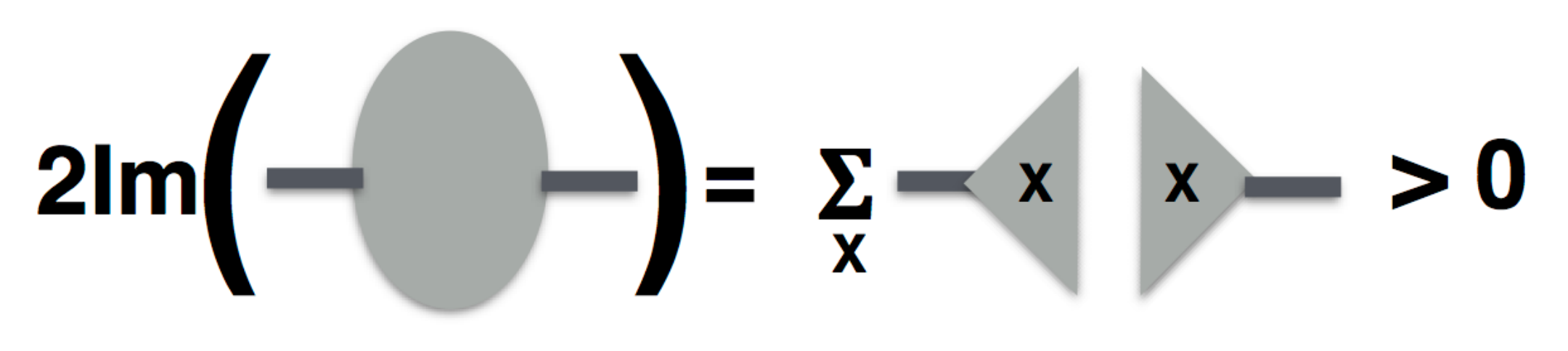} \; 
\end{center}
\caption{\small Standard sketch of optical theorem. The right-hand side is the sum over all intermediate states. 
It is a particular case of the cutting rules to be discussed  in section \ref{sec:CR}.}
\label{fig:optical}
\end{figure}

It is our goal to characterise the spectral function $\rho(s)$ in other ways than through the spectrum. 
In preparation to the general case we are going to  recite the optical theorem
for the S-matrix.
The S-matrix  \eqref{eq:Smatrix} is conveniently parameterised as 
\begin{equation}
S = \mathbb{1} + i T \;,
\end{equation}
where $T$ is the non trivial part of the scattering operator. From the unitarity of the S-matrix 
it follows that 
\begin{equation}
\mathbb{1} = S S^\dagger = \mathbb{1} +  \underbrace{i (T - T^\dagger)}_{- 2 {\rm Im}[T]} + |T|^2 \;,
\end{equation}
if and only if 
\begin{tcolorbox}
\begin{equation}
\label{eq:optical}
 2 {\rm Im}[T] = |T|^2 = T^\dagger \sum_n \state{n}\astate{n}  T   \;.
\end{equation}
Eq.~\eqref{eq:optical} is the celebrated \emph{the optical theorem} depicted in Fig.~\ref{fig:optical}. 
\end{tcolorbox}
The right-hand side of the equation above
is reminiscent of the spectral function in the case where $T$ is associated with $\phi$. Hence the expectation that $\rho(s)$ is related to an imaginary part is not unexpected from the viewpoint of the optical theorem. Below we are going to see that this is the case on very general grounds.

To do so we first take a little detour to discuss integral representations of arbitrary analytic functions 
by the use of Cauchy's theorem.
 Let  $f(p^2)$ be an analytic function then by Cauchy's theorem 
the following integral representation holds
\begin{equation}
f(p^2) = \frac{1}{2 \pi i} \int_\ga \frac{ds \, f(s) }{s-p^2} \;,
\end{equation}
provided that i) $p^2$ is inside the contour of $\ga$ , ii) the contour of $\ga$ does not cross 
any singularities. 

\begin{figure}
\begin{center}
\includegraphics[width=2.6in]{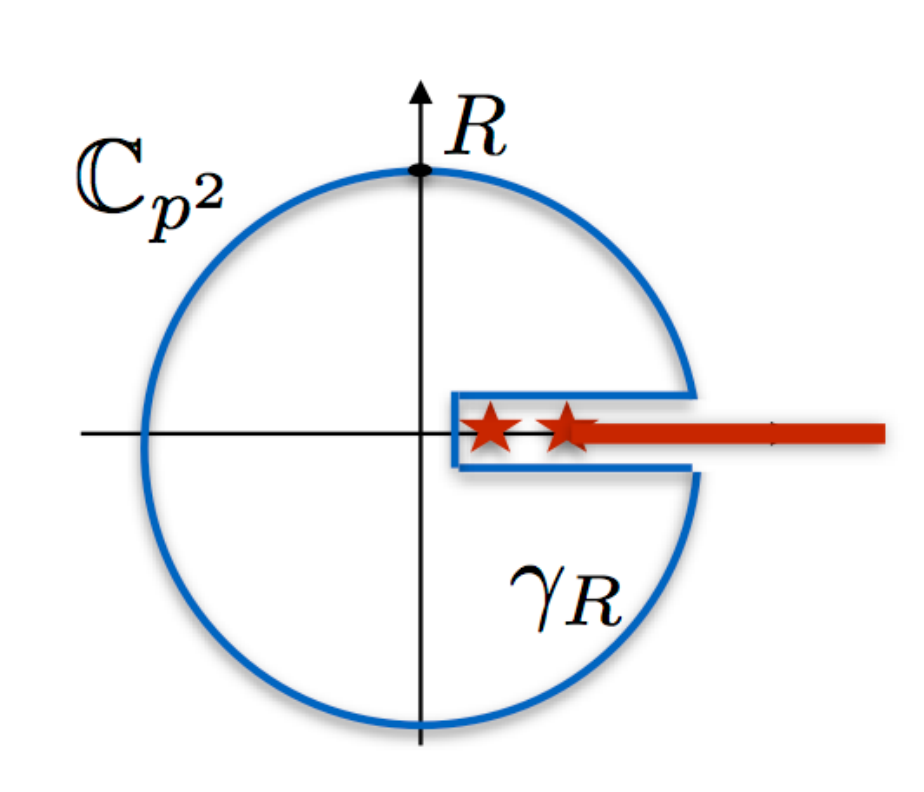} \; 
\includegraphics[width=2.0in]{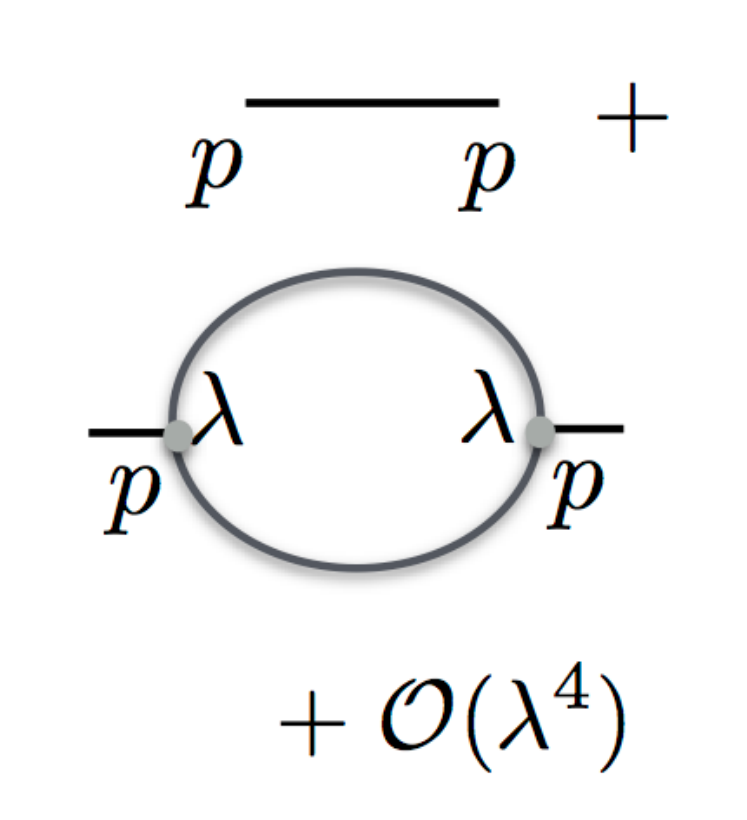}
\end{center}
\caption{\small (left) Integration contour  for 2-point function dispersion representation. (right) 
2-point function in $\phi^3$ theory in perturbation theory.}
\label{fig:pacman}
\end{figure}

Applying this techniques to the 2-point function in QFT one makes use of the knowledge 
of the analytic structure and chooses a contour $\ga_R$ 
as in Fig.~\ref{fig:pacman} which does not cross any singularities.  
The radius is then taken to infinity, $R \to \infty$, which in 
the case where there are UV divergences results in 
subtraction which we  generically parameterised $P(p^2)$ by a polynomial function 
(the $P = A$ in \eqref{eq:subtract} is a constant polynomial). 
The integral is then written as 
\begin{eqnarray}
\label{eq:disp}
\Gamma(p^2) &=& \frac{1}{2 \pi i} \int_{\ga_R} \frac{ds \, \Gamma(s) }{s-p^2} 
  \stackrel{R \to \infty}{\to}  \frac{1}{2 \pi i} \int_{s_1}^\infty \frac{ds \, ( \Gamma(s+i 0) - \Gamma(s-i 0)) }{s-p^2}     + P(p^2)  \nonumber \\[0.1cm]
&=&   \frac{1}{2 \pi i} \int_{s_1}^\infty \frac{ds \,  {\rm disc}[ \Gamma(s)] }{s-p^2 - i 0} + P(p^2) =     
 \frac{1}{ \pi } \int_{s_1}^\infty \frac{ds \, {\rm Im}[ \Gamma(s)] }{s-p^2 - i 0}  + P(p^2)    \;,
\end{eqnarray}
where $s_1 < \text{singularities}$ (to the left of red part in Fig.~\ref{fig:pacman}(left)) and
the second line is the definition of what is called the discontinuity along the branch cut. 
The last equality follows from $2i  {\rm Im}[ \Gamma(s)] =  {\rm disc}[ \Gamma(s)]$.
This formula can be verified in each order in PT but  is also  justified on general grounds
by the  Schwartz reflection principle (cf. appendix \ref{app:Schwartz}).
In summary we then have that the spectral function is related to the the imaginary part and the discontinuity  by
\begin{tcolorbox}
\begin{equation}
\label{eq:main}
\rho(s) = \frac{1}{\pi} {\rm Im}[ \Gamma(s)] = \frac{1}{2 \pi  i} {\rm disc}[ \Gamma(s)] \;.
\end{equation}
\end{tcolorbox}
This equation follows from equating \eqref{eq:KL} and \eqref{eq:disp} and the knowledge that the subtraction 
constant are the same in both cases since they originate from UV divergences. Hence 
eliminating the contributions from the arc may result in UV-divergences and subtraction constants.

\subsection{Dispersion Relations in Perturbation Theory}
\label{sec:disp-PT}

This section aims to illustrate \eqref{eq:main} from the viewpoint of PT.  In order to do PT one needs to 
specify a theory for which we choose  ${\cal L}_{\rm int} = \la \phi^3 + {\rm h.c.}$. The 
pole contribution is then just the propagator and the first non-trivial interaction is generated by 
the diagram in Fig.~\ref{fig:pacman}(right). 
The 1-loop graph is UV divergent and requires regularisation. Using 
dimensional regularisation $d = 4 - 2 \eps$ the result reads
\begin{equation}
\label{eq:above}
\Gamma(p^2) =  \frac{ Z(\la)}{ m^2 - p^2 -  i 0} -   \la^2 |A|\left(
\frac{1}{\eps} + 2  - \be \ln \left(  \frac{\be +1}{\be-1} \right) \right) + {\cal O}(\la^4) \;,
\end{equation}
with $\be = \sqrt{1 - (4m^2-i0)/p^2}$. 
The corresponding imaginary part divided by $\pi$ must be the spectral function  
\begin{equation}
\label{eq:rho}
\rho(p^2) \stackrel{\eqref{eq:main}}{=} \frac{1}{\pi} {\rm Im}[ \Gamma(p^2)]  
\stackrel{\eqref{eq:above}}{=} Z(\la)  \delta(p^2 - m^2)  + \la^2 |A| \be \theta(p^2 - 4 m^2)  + {\cal O}(\la^4) \;.
\end{equation}
The UV divergence is not important for our purposes 
since it does not affect the imaginary part. 
 Of course  the UV-divergence means that the dispersion relation 
does not converge in the UV. 
This problem can be handled either by a subtracted dispersion relation or a polynomial counterterm and regularisation as discussed under point 6 in section \ref{sec:KL}.

Having resolved this technical issue we focus on the interpretation of the imaginary part. 
The propagator term is a pole singularity with a delta function in the spectral function and the 
logarithm corresponds to a branch cut singularity resulting in a $\theta$-function part. 
By the spectral representation \eqref{eq:spec} this branch cut must correspond to some physical intermediate
state. This state is a 2-particle state starting at the minimum centre of mass energy $4m^2$ ranging all 
the way up to infinity. The precise value depends on the corresponding momentum configuration. Let the two particle momenta be parameterised by 
\begin{equation}
p_{1,2} = ( \sqrt{m^2 + x^2},0,0,\pm x)  \;, \quad x > 0 \;,   \quad p_{1,2}^2 = m^2 \;, \quad  p^2 = (p_1+p_2)^2 =  4 m^2 + 4 x^2
\end{equation}
and therefore $4 x^2 = p^2 - 4m^2$ which can be satisfied  for any (arbitrarily large)   $p^2 \geq 4 m^2$.

\section{Application of $2$-point Functions}
\label{sec:app}

There are numerous applications of $2$-point functions and dispersion relations. 
For example deep-inelastic scattering, QCD sum rules which we have alluded to in and below \eqref{eq:NP}, 
$e^+ e^- \to \text{hadrons}$ and
inclusive $b \to X_{u,c} \ell \nu$ decays with the additional assumption of analytic continuation to 
Minkowski-space.\footnote{Without going into any details let us mention that 
it is in particular the inclusive decay rate and amplitudes of exclusive decays that are amenable to 
a dispersive treatment. It is the amplitude and not the rate that has the simple analytic properties. 
The inclusive case is special in that the rate can be written as an amplitude!}
 We choose to present the Weinberg sum rules (WSR).

\subsection{Weinberg Sum Rules}

The Weinberg sum rules  are an ingenious construction involving a variety of conceptual ideas. 
They were proposed in 1967 by Weinberg \cite{WSR} in the pre-QCD era but we are going 
to present them from the viewpoint of QCD e.g. \cite{deRafael,Weinberg2}. 
One considers the correlation function of left and right-handed current with two massless quark flavours 
\begin{equation}
\label{eq:start}
i \int d^4 x e^{i q \cdot x} \vev{T J^{a,L}_\mu(x) J^{b,R}_\nu(x)}  = (q_\mu q_\nu - q^2 g_{\mu \nu} )
\Pi^{a,b}_{\rm LR}(q^2) \;,
\end{equation}
where 
\begin{equation}
J^{a,(L,R)}_\mu =   \bar q T^a \ga_\mu  q_{L,R} \;, 
\end{equation}
with $q_{L,R} = P_{L,R} q $, $P_{L,R} = 1/2 (1 \mp \ga_5)$,  $T^a$ being an $SU(2)$-generator (Pauli-matrix).
The Lorentz-decomposition in  \eqref{eq:start} is valid in the limit of massless 
quarks. According to the previous sections the  function $\Pi^{a,b}_{\rm LR}(-Q^2)$, with $-q^2 = Q^2 > 0$  satisfies a dispersion 
relation of the form 
\begin{equation}
\Pi^{a,b}_{\rm LR}(-Q^2) = \frac{1}{\pi} \int_{s_0}^\infty \frac{ds \, {\rm Im}[ \Pi^{a,b}_{\rm LR}(s)]}{s+Q^2} 
+ A
\end{equation}
where $A$ is a subtraction constant due to the potential logarithmic divergence which may arise since 
$\Pi^{a,b}_{\rm LR}$ is of mass dimension zero.

The peculiarity of the WSR relies on the absence of lower dimension  corrections 
in the OPE. This can be seen in an elegant manner using $SU(2)$ representation theory. We denote by $1,F$ and $A$ the trivial, fundamental and adjoint representation of $SU(2)$ which 
are of dimension $1,2$ and $3$.
The correlation function is in the 
(A,A)-representation of the $(SU(2)_L , SU(2)_R)$ global flavour symmetry. 
The individual OPE-contribution must  be  in the same  global flavour symmetry representation or vanish otherwise.

One considers Wilson's OPE in momentum 
space, valid for $Q^2 = - q^2 \gg \Lambda_{\rm QCD}^2$
\begin{equation}
\Pi^{a,b}_{\rm LR}(-Q^2) = C_{\mathbb{1}} (Q^2) \vev{\mathbb{1}} +  C_{\bar q q}(Q^2) 
\frac{ ( \vev{\bar q^{a}_L  q^{b}_R } + {\rm h.c.})}{Q^3}   +    C_{JJ} (Q^2) 
\frac{\vev{J^{L,a}_\mu J^{R,b}_\mu }}{Q^6}  + {\cal O}\left( \frac{\Lambda_{\rm QCD}^8}{Q^8} \right)  \;. 
\end{equation}
The functions $C(Q^2)$ are known as Wilson coefficients and carry logarithmic correction in QCD. 
As can be seen from the formula above  the condensate terms of dimension $d$ 
  are suppressed by $1/Q^d$ relative to the identity term.  
  The $\vev{\mathbb{1}}$-term corresponds to PT and the condensates, i.e. VEVs of operators, are of non-perturbative nature. The former  
  is in $(1,1)$-representation and  therefore absent.\footnote{The practitioner will notice the absence of 
   from the orthogonality of the projectors  $P_L P_R =0$ which necessarily arises in the a perturbative computation in the limit of massless quark.}
  A quark bilinear $\vev{\bar q^{a}_L  q^{b}_R }$ is in the $(F,F)$-representation and absent for the same reason. 
The dimension six operator is, somewhat trivially, in the $(A,A)$-representation and therefore the leading 
term appears at ${\cal O}(1/Q^6)$. The vanishing of the $1/Q^2$- and $1/Q^4$-terms 
 lead to constraints.
The latter can be obtained by expanding the denominator in inverse powers of $Q^2$,
\begin{equation}
\frac{1}{s+Q^2} = \frac{1}{Q^2}\frac{1}{1+s/Q^2} = \frac{1}{Q^2} - \frac{s}{Q^4} + \frac{s^2}{Q^6}  + \dots \;.
\end{equation}
\begin{tcolorbox}
The exact sum rules on the spectral function
\begin{eqnarray}
\label{eq:WSR}
  \int_{s_0}^\infty  ds \,   \Pi^{a,b}_{\rm LR}(s)   = 0 \;,  \qquad   \int_{s_0}^\infty  ds  \, \Pi^{a,b}_{\rm LR}(s)  s  = 0
\end{eqnarray}
known as the \emph{first and second Weinberg sum rule} follow. 
\end{tcolorbox}
Note the absence of the perturbative term means 
in particular that there is no UV-divergence and hence $A=0$. Since the convergence is 
two powers in $s$ higher  (1st and 2nd WSR) the dispersion relation is referred to a
as superconvergent. 

The WSR \eqref{eq:WSR} are a powerful non-perturbative constraint. We present the original application 
pursued by Weinberg.  First we notice that the left-right correlator can be written as a difference of the vector 
and axial correlator 
\begin{equation}
 \Pi^{a,b}_{\rm LR}(s) = \frac{1}{4}  \left(  \Pi^{a,b}_{\rm VV}(s) - \Pi^{a,b}_{\rm AA}(s)   \right) \;, 
\end{equation}
where $J^{V(A),a}_\mu \equiv \bar q T^a \ga_\mu ( \ga_5) q$.  
Taking into account the lowest lying particles $\pi$, $\rho$ and $a_1$ in the narrow width approximation and assuming isospin symmetry (i.e. global $SU(2)_V$ -flavour symmetry)
one arrives at\footnote{Note since we work in the massless limit 
the pion is massless as it is the goldstone boson of broken chiral symmetry $SU(2)_L \otimes 
SU(2)_R \to SU(2)_V$. The spin parity quantum numbers $J^P$ of the $\pi$, $\rho$ and $a_1$  are
$0^-,1^+,1^-$   respectively.}
\begin{alignat}{2}
\label{eq:spec}
& \rho_{V}^{a,b}(s) &\;=\;& \frac{1}{\pi} {\rm Im}[  \Pi^{a,b}_{\rm VV}](s) = \delta^{ab}  
(f_\rho^2 \de(s- m_{\rho}^2)  + \theta(s-s_0) \sigma_{V}) \;, \nonumber \\[0.1cm]
& \rho_{A}^{a,b}(s) &\;=\;& \frac{1}{\pi} {\rm Im}[  \Pi^{a,b}_{\rm AA}](s) = \delta^{ab}  
(f_\pi^2  \de(s) + f_{a_1}^2 \de(s- m_{a_1}^2)  + \theta(s-s_0) \sigma_{A}) \;.
\end{alignat}
The functions $\sigma_{V,A}$ contain any higher states and multiparticle states.  
If one assumes that around $s_0$ perturbation theory is valid then $ \rho_{\rm LR}(s) = 0$ for $s > s_0$ which 
in turn implies $\sigma_{V} = \sigma_{A}$.

Hence using \eqref{eq:spec}  the two WSR \eqref{eq:WSR} read
\begin{alignat}{2}
 f_\rho^2 =  f_\pi^2 +  f_{a_1}^2    \;, \qquad 
 m_\rho^2 f_\rho^2  = m_{a_1}^2 f_{a_1}^2 \;,
\end{alignat}
where the decay constants are defined as 
\begin{equation}
 \matel{\rho[a_1]^b(p)}{ J^{V[A],a}_\mu }{0}    = \de^{ab}  \eta_\mu(p) m_{\rho[a_1]} f_{\rho[a_1]}  \;, \quad  
  \matel{\pi^b(p)}{ J^{A,a}_\mu }{0}    = \de^{ab}  p_\mu  f_\pi \;,
\end{equation}
with $\eta$ being the polarisation vector.

In his original paper Weinberg used the experimentally motivated KSFR relation  $f_\rho^2  = 2 f_\pi^2$ which then leads 
to $m_{a_1} = \sqrt{2} m_\rho$.  This relation is reasonably satisfied by experiment: 
$m_{a_1}/m_\rho \simeq 1.63 \simeq 1.15 \sqrt{2}$.
Let us end this section with mentioning two further applications of this reasoning. 
\begin{itemize}
\item[-]
Being related to chirality 
the WSR, or the $\Pi_{LR}$ function, is a measure of contributions to electroweak precision measurement 
in the case of physics beyond the Standard Model coupling to new fermions. The WSR serve 
to estimate  the contribution
of  strongly coupled extensions of the standard model 
such as technicolor and the composite Higgs model.
\item[-] The inverse moments of the spectral function, with pion pole subtracted, are related to the low energy constant $L_{10}$ of chiral perturbation theory.  Note, chiral perturbation theory is an expansion in $Q^2$, and not 
$1/Q^2$ as the OPE, and thus leads to inverse moments rather than moments themselves. 
It is not the WSR per se which are important in this respect but the
onset of the duality threshold of PT-QCD which allows to estimate $L_{10}$ 
in terms of $f_{\pi,\rho,a_1}$. The estimate of $L_{10}$ obtained is in reasonable agreement with 
experiment. 
\end{itemize}

At this point the students were given the choice of continuing with applications 
(e.g. infrared interpretation of the chiral anomaly, positivity of low energy constants \dots)
or a the more conceptual topic of higher point functions properties.  Their  choice
was the latter. 

\section{Analyticity Properties of higher point Functions}
\label{sec:higher}

There are many motivations to study higher point functions and their analytic structure amongst which 
we quote the following: 
\begin{itemize}
\item[a)] As seen in the introduction they describe the scattering of $n$-particles. 
\item[b)] From the discussion in section \ref{sec:cauchy} it is clear that to write down dispersion relations
one needs to know first and foremost the analytic structure of the amplitude in question.
\item[b)] 3-point functions are relevant for the study of form factors. Consider for 
example the $B \to \pi$ form factor, relevant for the determination of the CKM-element $|V_{\rm ub}|$, 
defined by 
\begin{equation}
\matel{\pi(p)}{V_\mu}{B(p_B)} = (p_B)_\mu f^{B \to \pi}_+(q^2) +  \dots \;,
\end{equation}
where the dots stand for the other Lorentz structure and $V_\mu = \bar b \ga_\mu u$ is the weak current (the axial part does not contribute in QCD by parity 
conservation). Then the form factor can be extracted from the following 3-point function, 
by using a double dispersion relation (dispersion relation in the $p_B^2$ and $p^2$-variable) 
\begin{eqnarray}
\label{eq:double}
\Gamma(p^2,p_B^2,q^2) &\;=\;&  i^2 \int_{x,y} e^{ - i ( p_B \cdot x - p \cdot y)} \vev{T J_B(x) J_\pi(y) V_\mu(0)}  
\;, \nonumber \\[0.1cm] 
&\;=\;&    
(p_B)_\mu \left( \frac{g_\pi \,  f_B \,  f_+^{B \to \pi}(q^2)}{(p_B^2-m_B^2)(p^2-m_\pi^2)} + \text{higher} \right) + \dots \;,
\end{eqnarray}
where  ``higher" stands for higher contributions in the spectrum (the analogue of $\sigma(s)$ in \eqref{eq:rho}) 
and 
\begin{alignat}{4}
& J_B &\;=\;& \bar q i \ga_5 b   \;,   \quad  & &  \matel{B}{J_B}{0} &\;=\;& f_B  \;, \nonumber \\[0.1cm]
& J_\pi &\;=\;& \bar q i \ga_5 q   \;,   \quad & & \matel{0}{J_\pi}{\pi} &\;=\;& g_\pi \;, 
\end{alignat}
play the role of the interpolating operators of the LSZ-formalism cf. footnote \ref{foot:inter}.  
As previously mentioned the key idea is then to compute $\Gamma(p^2,p_B^2,q^2) $ in some formalism 
and to find ways to either estimate or suppress the higher states in order to extract the form factor 
where $g_\pi$ and $f_B$ are assumed to be known quantities. 

In fact if we were able to compute $\Gamma(p^2,p_B^2,q^2)$ with arbitrary precision then 
the function would assume the form in \eqref{eq:double} and we could simply 
extract the form factor from  the limiting expression   
\begin{equation}
f_+^{B \to \pi}(q^2) =  \frac{1}{g_\pi f_B}   \lim_{p_B^2 \to m_B^2, p^2 \to m_\pi^2} (p_B^2-m_B^2)(p^2-m_\pi^2)\Gamma(p^2,p_B^2,q^2) \;,
\end{equation}
which makes the connection to the LSZ-formalism \eqref{eq:LSZ} apparent.
Unfortunately at present we cannot hope to do so and therefore we have to resort to the approximate
techniques as alluded to above.
\end{itemize}

We have seen that for the 2-point function the  analytic structure of the first sheet (physical sheet)  is fully
understood through the K\"all\'en-Lehmann representation.  Moreover the singularities on the physical 
sheet are in one-to-one correspondence with the physical spectrum. For higher point function less 
is known in all generality. We refer the reader to the works of K\"all\'en-Wightman \cite{three} and  
K\"all\'en \cite{four} for some general studies of 3- and 4-point functions
using first principles and the summary by Andr\`e Martin \cite{Martin} for a comparatively recent survey 
of rigorous results.\footnote{Am important topic was the 
conjecture by Mandelstam of a double dispersion relation for $ 2 \to 2$ scattering (i.e. 4-point function) 
which was consistent  with known results but never proven in all generality even in perturbation theory. 
From this the  Froissart bound was derived which states that the cross section for the scattering of two particles
cannot grow faster than $\ln^2 s$ (where $s$ is the centre of mass energy).}
 
 Hence one has to become immediately more modest! We are
 going to  restrain ourselves  to analysing  singularities in PT for \emph{physical  momenta} (i.e. real momenta). 
This is done by the use of two major tools:
\begin{itemize}
\item [i)] \emph{Landau equations}: which answer the question about the location of  the singularities  cf. section \ref{sec:LE}. The question on which sheet the singularities are is 
a difficult question which we comment on.
\item [ii)] \emph{Cutkosky rules}: are rules for  computing the discontinuity of an amplitude cf. section \ref{sec:CR}.
\end{itemize}
Before analysing these matters in more details let us first consider the  normal-thresholds 
for higher point functions.

\subsection{Normal Thresholds: cutting  Diagrams into two Pieces}
\label{sec:normal}

The so-called normal thresholds are directly associated with unitarity. 
They originate from cutting (to be made more precise when discussing the Cutkosky rules) 
the diagram into two pieces and 
 generalise the equal size optical theorems cuts.
   Cutting a diagram into two pieces is equivalent to the combinatorial problem 
of grouping the external momenta into two sets. Tab.~\ref{tab:cuts} provides the overview of the number 
of cuts versus number of independent kinematic variables. For the two lowest functions there are no constraints 
whereas for all higher point functions there are constraints due to momentum conservation. For the $4$-point functions this constraint is known 
as the famous \emph{Mandelstam constraint}
\begin{equation}
\label{eq:MSc}
\sum_{i=1}^4 p_i^2 = s + t + u \;, \quad s = (p_1+p_2)^2 \;,  \quad  t = (p_1+p_3)^2 \;, 
\quad u = (p_1+p_4)^2 \;,
\end{equation}
where we choose conventions such that all momenta are incoming 
(momentum conservation reads $\sum_{i=1}^4 (p_i)_\mu =0$).
\begin{figure}[h]
\begin{center}
\includegraphics[width=4in]{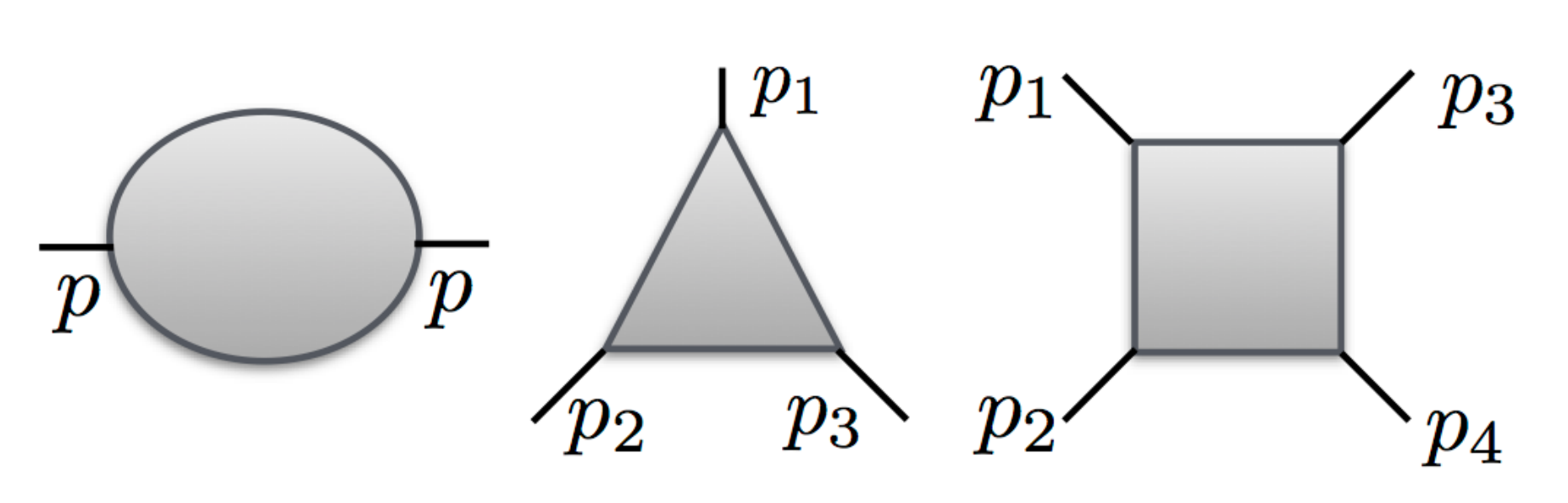} \; 
\end{center}
\caption{\small Sketch of generic $2,3,4$-point function. The number of unitarity cuts and independent kinematic variables are listed in Tab. \ref{tab:cuts}.}
\label{fig:three}
\end{figure}
The fact that the amplitudes become functions of several complex variables makes matters more difficult. For example:
\begin{itemize}
\item  From the Mandelstam constraint it can be seen that 
the unitarity cuts from one channel do appear on the negative real axis in 
the complex plane of another channel. Say the $u$-channel cuts, with forward kinematics 
$t=0$, do appear on the negative real axis in the complex plane of $s = 
\sum_i m_i^2 - u $ ($p_i^2  = m_i^2$ on shell). These $u$-channel 
cuts in the $s$-plane are sometimes referred to as left-hand cuts as opposed to the proper
$s$-channel unitarity cuts on the right-hand side of the complex plane (cf. Fig.~\ref{fig:pacman}). 
\item
Several complex variables allow for  dispersion relations in multiple channels. 
An example of which is the conjectured Mandelstam representation, cf. \cite{IZ,S-matrix},  which is a double dispersion relation.
\item There are cuts which are not directly related to unitarity, so-called anomalous thresholds, cutting the 
diagram into more than 2 pieces. We will return to the latter briefly when discussing the Landau equations and Cutkosky rules.
\end{itemize}
\begin{table}[h]
$$
\begin{array}{c  |  c | c | c   }
n\text{-point function} & \text{\#cuts} & \text{\#variables} & \text{\#constraints}  \\ \hline
2 & 1 & 1 & 0 \\
3 & 3 & 3 & 0 \\
4 & 7 & 6 & 1  \\
5 & 15 & 10 & 5 
\end{array}
$$
\caption{\small The number (=\#)  of unitarity cuts equals the number of independent variables plus the number of constraints. Cf.  Fig.~\ref{fig:three} some diagrams.
For 
the $2$- and $3$-point functions there are no constraints whereas for the $4$-point function there is the famous 
Mandelstam constraint \eqref{eq:MSc}. 
Generally for $n \geq 4$ the number of invariants is $ 4 n - 10$ where $4n$ corresponds to  all the components of the four momenta and $10$ subtracts the Lorentz-symmetries. (For $n \leq 4$ 
the formula reads $n(n-1)/2$ which gives a larger number since not all the Lorentz symmetries can be used for reduction of parameters). For the $5$-point function 
the reduction from 15 cuts to 10 independent variables 
follows from multiple use of the Mandelstam relation \eqref{eq:MSc} on subtopologies of the kinematic diagram e.g.\cite{kinematics}.}
\label{tab:cuts}
\end{table}

\subsection{Landau Equations}
\label{sec:LE}

Before stating the Landau equations it is useful to look at singularities of a one-variable 
integral representation where the integrand has pole singularities as a function of external 
parameters. The Landau equations originate from analysing this problem for the integrals 
of several variables appearing in PT. The presentation in this section closely follows
the original paper \cite{Landau} and the textbook\cite{S-matrix}.

\subsubsection{Singularities of one-variable Integral Representations}
\label{sec:singular}

\begin{figure}
\begin{center}
\includegraphics[width=6in]{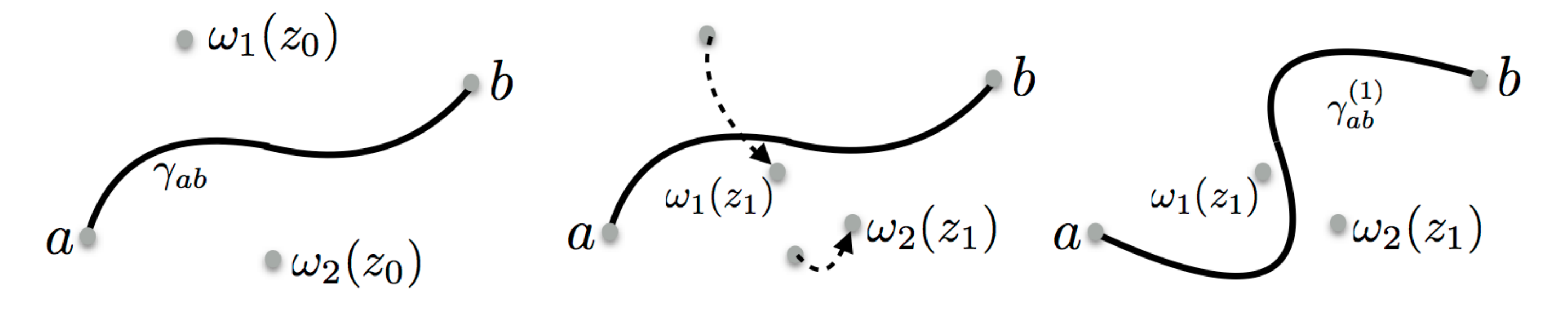} 
\end{center}
\caption{\small Path $\ga_{ab}$ between endpoints $a$ and $b$. 
(left) Poles $\omega_{1,2}(z_0)$ of the integrand $g(z,\omega)$ in \eqref{eq:fz} for  $z=z_0$ (middle) Deformation of poles $\omega_{1,2}$  by the parameter $z_0$ to $z_1$. 
This is not provide an analytic continuation of the function $f(z)$ around $z_0$ to $z_1$.
(right) deformation of path $\ga_{ab}$ to $\ga_{ab}^{(1)}$ serves as a (legitimate) analytic continuation of the function $f(z)$ in \eqref{eq:fz} around $z_0$ to $z_1$.}
\label{fig:oneC}
\end{figure}

Consider the following integral representation of a analytic function $f(z)$
\begin{equation}
\label{eq:fz}
f(z) = \int_{\ga_{ab}}  g(z,w) dw \;,
\end{equation}
where the integrand $g(z,w)$ contains pole singularities $w_i(z)$ which depend on $z$.   
The path $\ga_{ab}$ ranges from a point $a$ to $b$ and does not cross any singularities for some $z= z_0$
as shown in  Fig.~\ref{fig:oneC}(left).  The  analytic properties of $f(z)$  depend 
on whether or not the path $\ga_{ab}$ can be smoothly deformed away from 
approaching pole singularities $\omega(z_i)$.

\begin{figure}
\begin{center}
\includegraphics[width=5in]{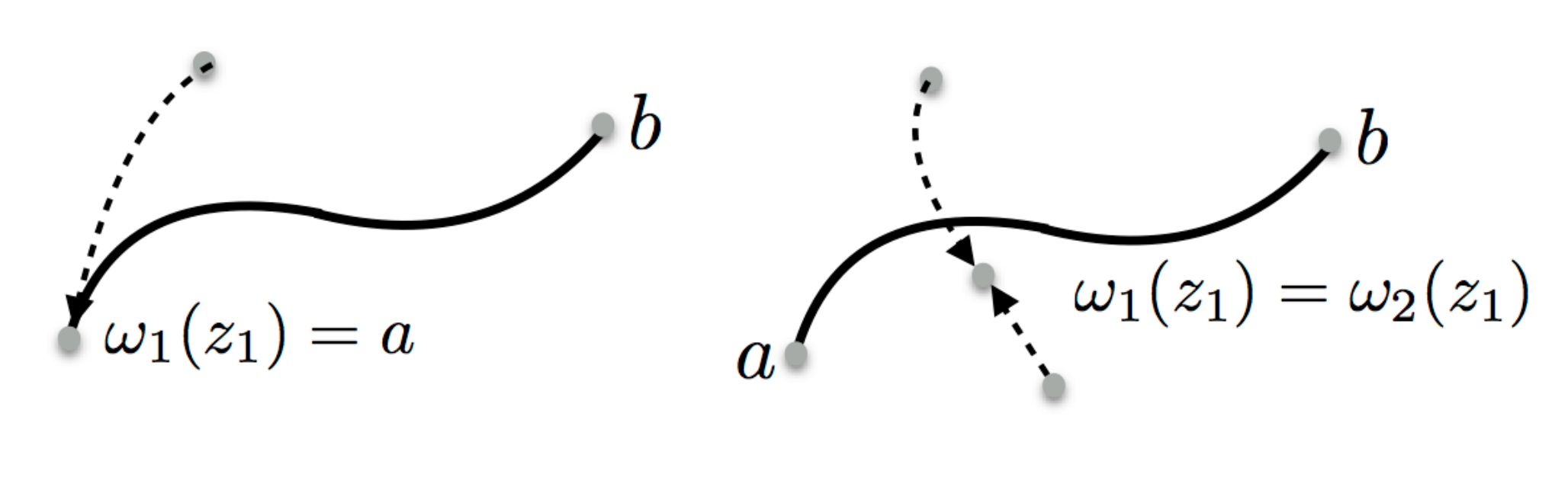}
\end{center}
\caption{\small (left) endpoint singularity  (right) 
pinch singularity}
\label{fig:sing}
\end{figure}

For example, if we start from $z=z_0$ and go to $z=z_1$ with $\omega(z_1)$ crossing
the   $\ga_{ab}$ (cf.  Fig.~\ref{fig:oneC}(middle)) then the path $\ga_{ab}$ can be smoothly deformed 
as in  Fig.~\ref{fig:oneC}(right) and this constitutes an analytic continuation of the function $f(z)$.
There are though instances when this is not possible:
\begin{itemize}
\item[a)] When a singularity $w_i(z)$ approaches  one of the endpoints  $a$ or $b$; e.g. $w_i(z_1) = a$ 
Fig.~\ref{fig:sing}(left).
This case is known as an \emph{endpoint singularity}.
\item[b)] When two singularities approach each other, $w_1(z_1) = w_2(z_1)$ from different direction of the integration path as depicted in 
Fig.~\ref{fig:sing}(right). This case is known as a \emph{pinch singularity}.
\item[c)] When the path needs to be deformed to infinity (can be reduced to case b).
\end{itemize}
In PT it is the pinch singularity type that gives rise to the singularities.

\subsubsection{Landau Equations = several-variable Case}

\begin{figure}
\begin{center}
\includegraphics[width=2.5in]{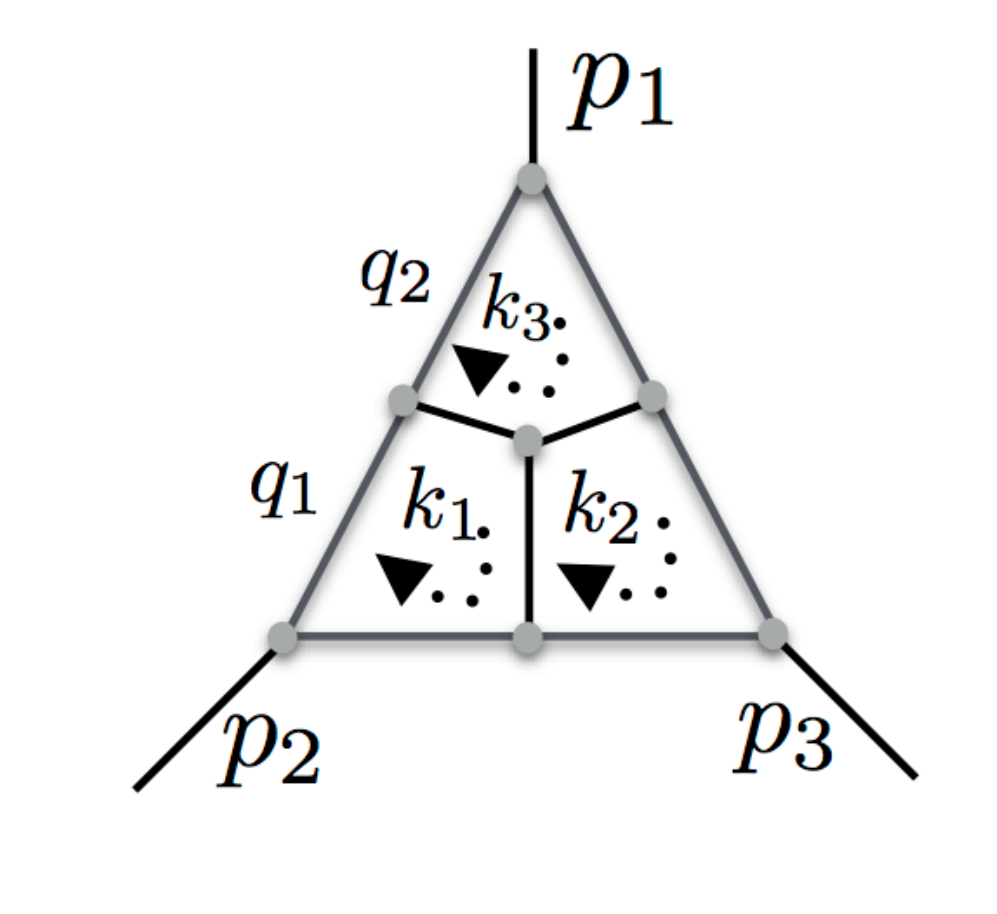} \; 
\includegraphics[width=2.2in]{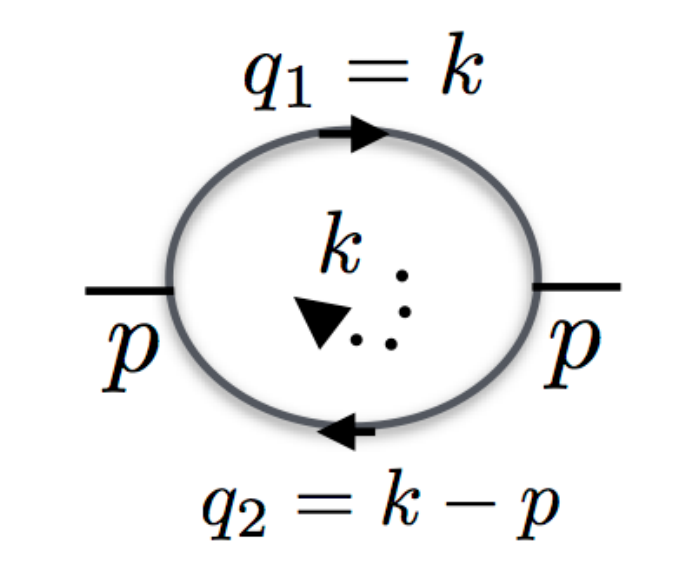}
\end{center}
\caption{\small (left) Generic Feynman diagram aimed to clarify notation used in text 
(right) bubble graph discussed in the text}
\label{fig:rep}
\end{figure}

Landau \cite{Landau} and others (cf. \cite{S-matrix} for further references ) have analysed the problem of
singularities, discussed for a single integral above, for the case of several variables 
in the context of Feynman graphs.   
A generic Feynman graph of $L$-loop of momenta $k_i$ ($i=1 \dots L$), $N$-propagators, external momenta $p_i$
(cf. Fig.~\ref{fig:rep}(left) for a  representative graph) can be written as follows
\begin{equation}
\label{eq:I}
I = \int Dk \frac{1}{\prod_{i=1}^N (q_i^2 -m_i^2 + i 0)} \;,  \qquad Dk  = \prod_{i=1}^L d^4 k_i \;,
\end{equation}
where $q_i = q_i(p_j,k_l)$ are the momenta of the propagators. By the technique of Feynman parameters
(generalisation of $(A B)^{-1} = \int_0^1 d\al (\al A + (1-\al) B)^{-2}$) one may rewrite $I$ as follows
 \begin{equation}
 \label{eq:D}
I = \int Dk  \int_0^1 D\al  \frac{1}{(F+ i0)^N} \;,  \qquad D \al  = \prod_{i=1}^N d \al_i \de(1 - \sum_{i=1}^N \al_i) \;,
\end{equation}
where the crucial denominator $F$ reads
\begin{equation}
\label{eq:F}
F = \sum_{i=1}^N \al_i (q_i^2 - m_i^2 ) \;.
\end{equation}
Is seems worthwhile to emphasise that even though these formulae look rather involved they are completely straightforward.

The key idea is that there are different types of singularities depending on how many of the 
$N$ propagators are on shell, i.e. $q_i^2 - m_i^2$.  It is the number of on-shell propagators 
which serves as a classification of the singularities. 
The Landau equations in condensed form are:\footnote{The Landau equations enjoy an interpretation in terms of electric circuits since the equations are analoguous to Kirchoff's equations cf. \cite{S-matrix} and references therein.}

\begin{tcolorbox}
\paragraph{Landau equations/conditions} There are singularities if and only if
\begin{eqnarray}
\label{eq:LE1}
& &  \text{[i]      either } q_i^2 = m_i^2 \text{ or } \al_i = 0 \;,  \\[0.1cm]
\label{eq:LE2}
& &  \text{[ii] } \sum_{i \in \text{loop}(l)}   \al_i (q_i)^\mu =0 \text{ for } l=1  \dots L.
\end{eqnarray}
We have assumed $q_i = k + \dots$ with $k$ being the loop momentum as otherwise one needs to introduce spurious minus signs into the sum above.
\end{tcolorbox}
We are not going to show a proof of the Landau equations 
but state the result and argue for its plausibility below.
Let us emphasise that the Landau equations neither tell us on which sheet the singularities are (cf. section \ref{sec:example} for the refinement in this direction) nor how to compute 
the discontinuity relevant for the dispersion relations (cf. Cutkosky rules  section \ref{sec:CR}).
The first condition assures that $F=0$ by demanding that each summand is zero in \eqref{eq:F}. 
The interpretation of $q_i^2 = m_i^2$ is of course that the corresponding propagator is on-shell 
and contributes to the singularity. Correspondingly $\al_i = 0$ means that the corresponding line does not 
enter the singularity.  In Fig.~\ref{fig:three-cuts} we give an example of a 3-point function cut.
The second condition \eqref{eq:LE2}  has a geometric interpretation. It means that the corresponding singularity surfaces 
are parallel to each other and that the hypercontour can therefore not be deformed away from the approaching 
singularity surfaces. This is the analogy of the pinch singularity discussed in section \ref{sec:singular}.
Eq.~\eqref{eq:LE2} can be cast into a more convenient form by contracting the equation by a vector 
$(q_j)_\mu$ which leads to the matrix equatons
\begin{equation}
\label{eq:LE2b}
\text{[ii']} \quad  Q \vec{\al} = 0 \;,     \qquad (Q)_{ij} = q_i \cdot q_j \;, \quad  (\vec{\al})_i =  \al_i \;.
\end{equation}
For non-trivial $\vec{\al}$ the second Landau equation is then solved by demanding 
that the so-called Cayley-determinant $\rm det Q =0$ vanishes.

\subsubsection{Landau Equation exemplified: 1-loop 2-point Function (bubble graph)}
\label{sec:example}

Consider the bubble graph depicted in Fig.~\ref{fig:rep}(right) with external momenta $p$, loop momenta $k$ 
and momenta $q_1 = k$ and $q_2 = k-p$. 
The first Landau equation \eqref{eq:LE1} tells us that [As a homework you could ask yourself why the case 
$\al_1 = 0$ and $q_2^2 = m_2^2$ is not an option for a singularity] $q_1^2 = m_1^2$ and $q_2^2 = m_2^2$.
 The second Landau equation \eqref{eq:LE2b} can be cast into the form $\det Q =0$
\begin{equation}
\det Q = \det   \left(  \begin{matrix}   m_1^2  & q_1 \cdot q_2  \\  q_1 \cdot q_2 & m_2^2  \end{matrix} \right) 
= 0  \quad \Leftrightarrow q_1 \cdot q_2  = \pm m_1 m_2
\end{equation}
which we may reinsert back into $p= q_1 + q_2$ which yields the two singularities 
$p_{(+)}^2 = (m_1+m_2)^2$ \emph{and}  
$p_{(-)}^2 = (m_1-m_2)^2$,
\begin{equation}
p^2 = (q_1 - q_2)^2 = q_1^2 - 2 q_1 \cdot q_2 + q_2^2 = (m_1 \mp m_2)^2 \;.
\end{equation}
 This might surprise us at first since from unitarity we expect there to 
be a branch point at $p_{(+)}^2$ but the point $p_{(-)}^2 < p_{(+)}^2$ has no place in this picture. 
The resolution comes upon recalling that the Landau equations inform us about the singularities but 
do not tell us on which sheet they are! In order to learn more we may solve 
$Q \vec{\al} = 0$ with $\vec{\al} = (\al , (1-\al))^T$ which gives 
\begin{equation}
\al_{\pm} = \frac{m_2}{m_2 \pm m_1}  \Rightarrow  \quad  0 < \al_+ < 1 \;,  \quad  \al_- > 1 \text{ or } \al_- <0  
\;,
\end{equation} 
for $m_{1,2} > 0$.
From this we learn that $\al_+$ is within the integration region (recall $\int_0^1 D\al$ 
\eqref{eq:D}) and $p_{(+)}^2$ is therefore on the physical sheet,  
wheras $\al_-$ is outside the integration region necessitating the deformation of  the $\al$-contour.
This indicates that $\al_-$ may not be on the physical sheet as the contour
may crosses singularities in the course of deformation.  The singularity $p_{(-)}^2  = 
(m_1-m_2)^2$ is sometimes referred to as a pseudo threshold.
 These findings suggest  an important refinement 
of the Landau conditions.
\begin{tcolorbox}
\paragraph{Refinement of Landau equations}
For physical configuration, by which we mean the real external momenta, the Landau singularities are 
\begin{itemize}
\item[-] on the first (physical) sheet when $\al_i \in [0,1]$
\item[-] may or may not be on the first (physicial) sheet  when $\al_i \notin [0,1]$
\end{itemize}
\end{tcolorbox}

For non physical configuration, complex momenta, the situation is far from straightforward to say the least. 
The method of choice is often deformation to a case of a physical configuration and then deform to complex momenta  
checking whether or not singularities are crossed  in that process.   
Crossing a singularity correspond to changing the Riemann sheet.
Alternatively one can deform the masses
to complex values keeping the $\al_i \in [0,1]$ and then deform back.

\paragraph{Geometric interpretation and the forgotten second type singularity}

\begin{figure}
\begin{center}
\includegraphics[width=2.0in]{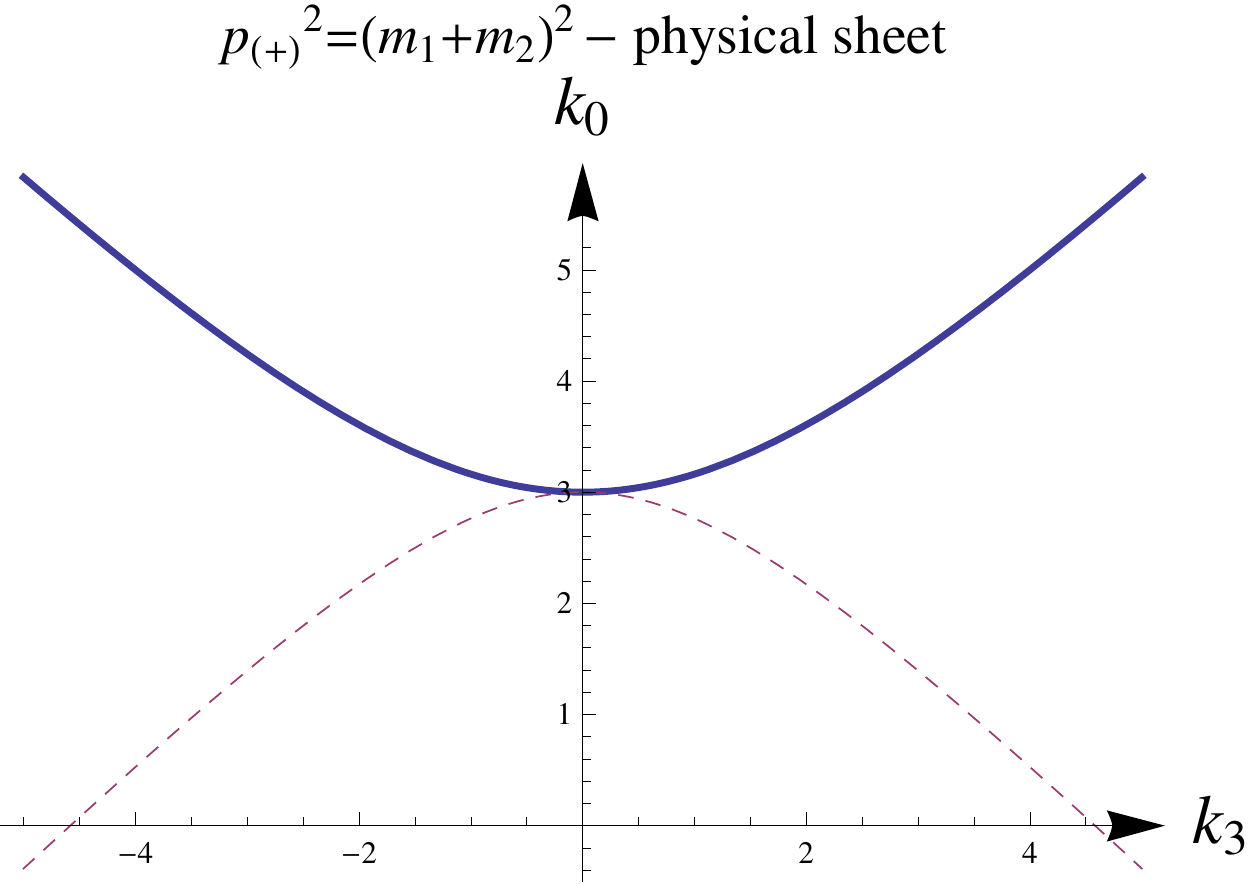}  \includegraphics[width=2.0in]{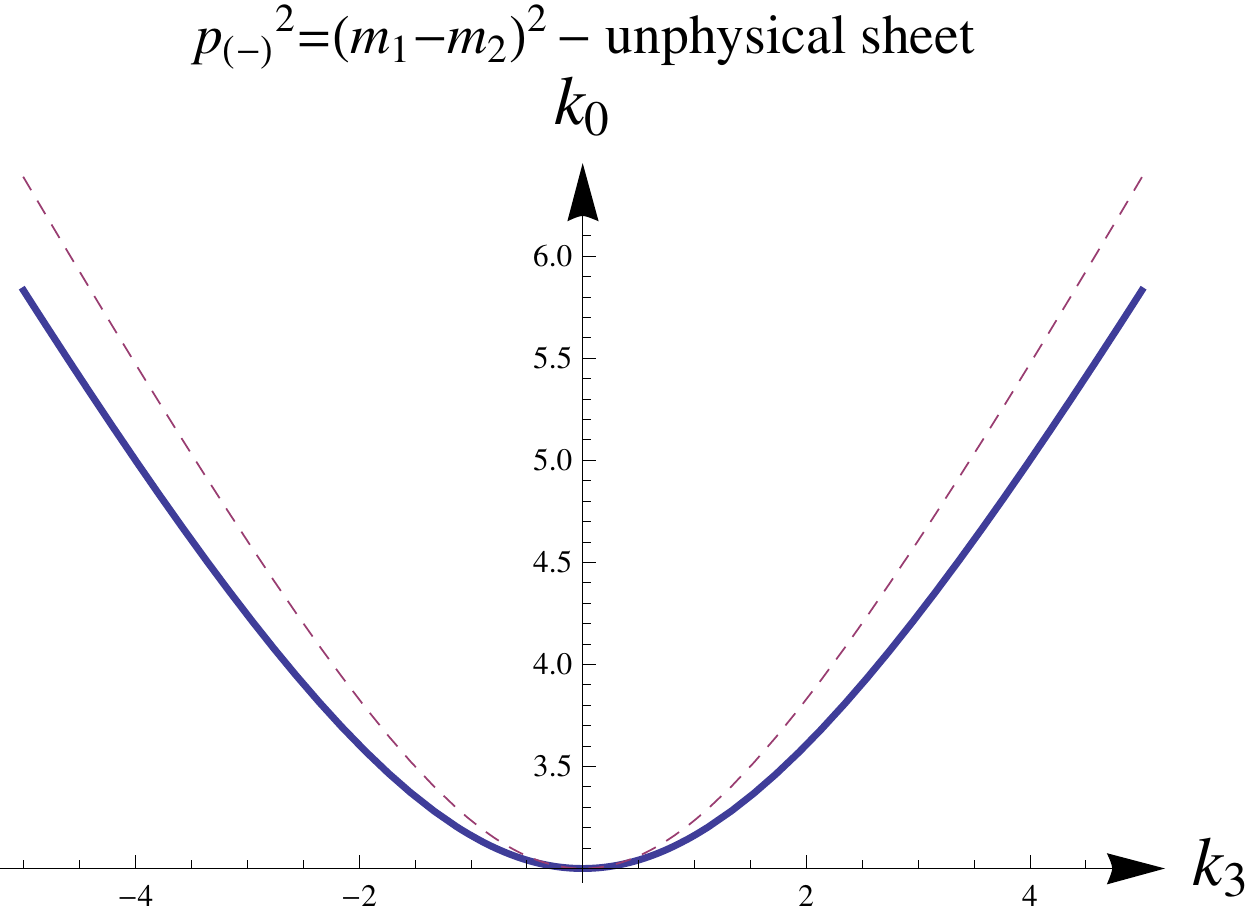}  \includegraphics[width=2.0in]{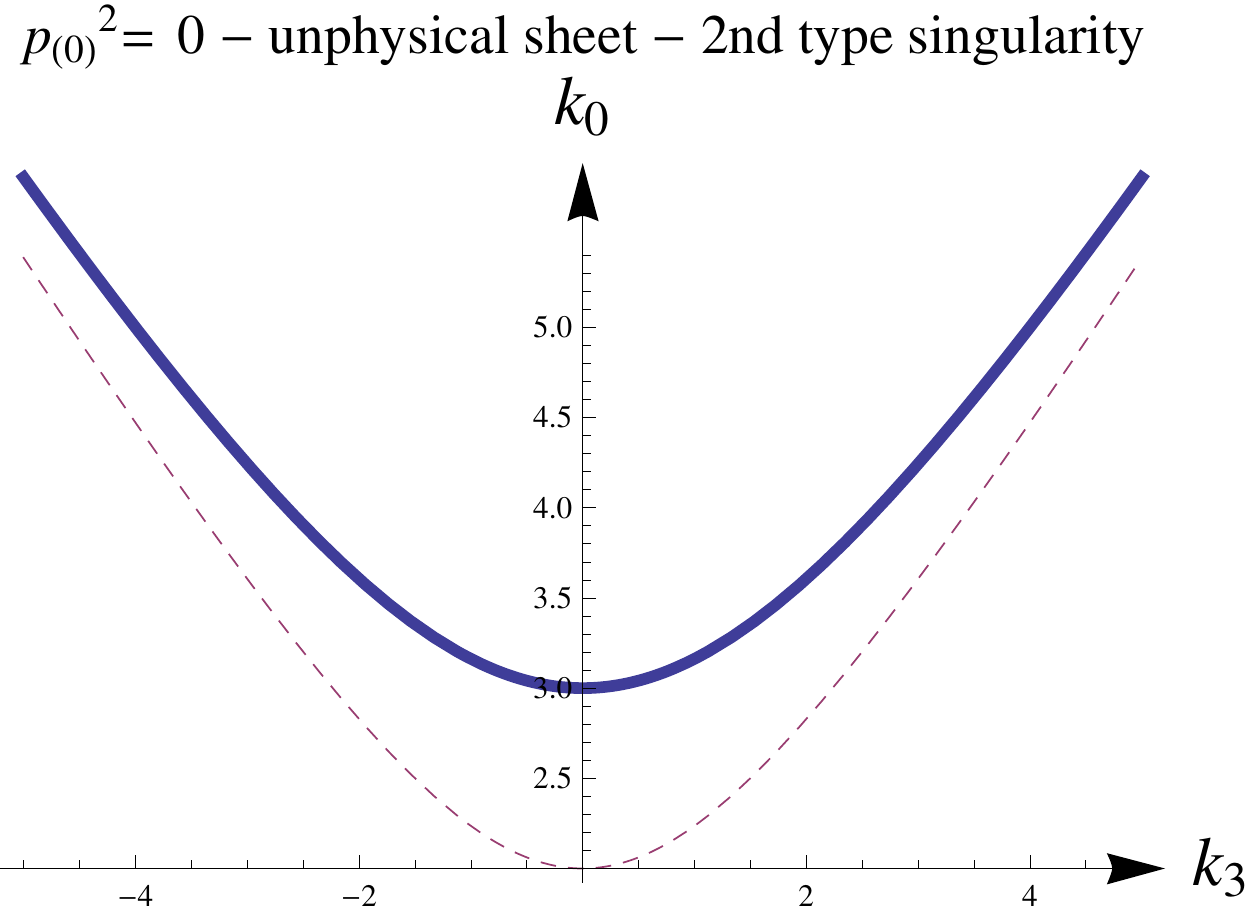}
\end{center}
\caption{\small Choosen $m_1= 3$ and $m_2 = 2$ in arbitrary units. 
(left) \emph{Normal threshold:} $k_0 =  \sqrt{k_3^2 + m_1^2}$ and $k_0  = (m_1+m_2)  - \sqrt{k_3^2 +m_2^2}$  and $p_+ = (m_1+m_2,0,0,0)$. (middle) 
\emph{Pseudo threshold:} $k_0 =  \sqrt{k_3^2 + m_1^2}$ and $k_0  = (m_1-m_2)  + \sqrt{k_3^2 +m_2^2}$  and $p_- = (m_1-m_2,0,0,0)$.  (right) 
\emph{Second type singularity:}
$k_0 =  \sqrt{k_3^2 + m_1^2}$ and $k_0  =  \sqrt{k_3^2 +m_2^2}$  and $p_{(0)}  = (0,0,0,0)$  the two curves meet at infinity and are independent of the masses. 
It is obvious that any light-like $p$, e.g. $p_{(0)} = (x,0,0,x)$ for $x$ real, is also a valid parametrisation.}
\label{fig:hyper}
\end{figure}

We take a detour and give a geometrical interpretation of the singularities
of the bubble graph.
 The first and second Landau equations (\ref{eq:LE1},\ref{eq:LE2}) read  
 \begin{equation}
 \label{eq:LEB}
 (k-p)^2 = m_1^2 \;, \quad k^2 = m_2^2 \;, \quad 
 \al k_\mu  + (1-\al) (k-p)_\mu = 0  \;.
 \end{equation}  The first Landau equation defines two hyperboloids with centres displaced by $p$ with respect to each other.
The second Landau equation assures that $p$ and $k$ are parallel.  
First we discuss the previously found solutions and then uncover the forgotten 
second  type singularity.
\begin{itemize} 
\item \emph{Normal and pseudo threshold} \\
Equations \eqref{eq:LEB} are satisfied when the two hyperboloids touch each other at the symmetric point with displacement in the $k_0$-direction.  
The displacement is given by $p_{(\pm)}^2 = (m_1 \pm m_2,0,0,0)^2 = (m_1\pm m_2)^2$ in the case where the hyperboloids open up in opposite and the same 
direction respectively. We refer the reader to  Fig.~\ref{fig:hyper} (left) and (middle) for an illustration and the relevant equations in the caption. 

\item \emph{Second type singularity}  \\
Fig.~\ref{fig:hyper} (right) shows yet another type of singularity. For any light-like displacement 
$p$, the two hyperboloids meet at infinity. These type of singularities were  first noted by Cutkosky \cite{CR} who named them 
non-Landauian singularities whereas nowadays they are known as \emph{second type 
singularities} \cite{S-matrix}. The Cayley-determinant is zero since $q_1 $ and $q_2 $ 
are parallel with $k$ and $p$ being light-like. If we parameterise $p/k=\eps$ then 
the second Landau equation is satisfied when $ \al = (1+\eps)/\eps$ which diverges 
when $\eps \to 0$ and in particular outside the $[0,1]$-interval. 
Hence  the singularity may therefore not be on the physical sheet.  More directly  the singularity $p_{(0)}^2 = 0$ has no interpretation in terms of the spectrum.  By the one-to-one relation of the spectrum and the 
2-point function  singularities (cf. K\"all\'en-Lehmann representation in section \ref{sec:2pt}) it is to be concluded that  this singularity is not on the physical sheet. 
\end{itemize} 
 
 More  generally  second type singularities are determined by the vanishing of 
the Gram-determinant $\det p_i \cdot p_j = 0 $ (where the $p_i$ are the external momenta)  \cite{S-matrix}. 
The singularity $p_{(0)}^2 = 0$ for the bubble-graph is then just the solution of the one-dimensional Gram-determinant. 
 In summary second-type singularities are therefore independent of the masses and not on the physical sheet.\footnote{There also exist mixed type singularities where some loop momenta are pinched at infinity and others not. The reader is referred to the book \cite{S-matrix} for examples and discussion.} 

\paragraph{Discussion upon using explicit 1-loop result:}

After these abstract considerations we consider it advantageous to illustrate these 
type of singularities explicitly. The bubble graph 
\begin{equation}
B_0(p^2, m_1^2,m_2^2) = 
 \frac{ 4 \pi^2}{ i}\int \frac{d^4 k}{(2 \pi)^4 } \frac{1}{(k-p)^2 - m_1^2+i0)(k^2 - m_2^2+i0)}  \;,
\end{equation}
is UV-divergent. In order to avoid regularisation we may use the same trick 
as for the subtracted dispersion relation and take the difference with a fixed value. 
The results, valid on the physical sheet, reads (e.g. \cite{IZ})
\begin{equation}
\label{eq:explicit}
B_0(p^2,m_1^2,m_2^2)  - B_0(\bar{p}^2,m_1^2,m_2^2)  = 
\frac{1}{2 p^2}\left[ X(p^2,m_1,m_2) -X(0,m_1,m_2) \right]    - [ p^2 \leftrightarrow \bar{p}^2] 
\;,
\end{equation}
where $\bar{p}^2$ is the arbitrary subtraction point, 
\begin{equation}
X(p^2,m_1,m_2) =   \sqrt{\la} \ln \left( \frac{ \sqrt{(m_1+m_2)^2 -p^2 } + 
\sqrt{(m_1-m_2)^2 -p^2 }    }{ \sqrt{(m_1+m_2)^2 -p^2 } - 
\sqrt{(m_1-m_2)^2 -p^2 }   }  \right)^2 
\end{equation}
and $\la =  ( p^2- (m_1+m_2)^2) (p^2- (m_1-m_2)^2) $
the K\"all\'en-function and not to be confused with a coupling constant.\footnote{From the viewpoint of the optical theorem 
the K\"all\'en-function arises from the phase space integration. 
In the rest-frame of the  particle associated with the four momentum squared $p^2$, 
the absolute velocity of one of the decaying particles $|v_1| = |v_2|$
is related to the K\"all\'en-function as follows $|v_{1,2}| = \sqrt{\la}/2p^2$. 
Hence at this singularity the  velocity is infinite consistent with the hyperboloids meeting at infinity.}  It is noted that the expression above is consistent with \eqref{eq:above} in the limit $m= m_1 =m_2$.
We can learn three things from the representation \eqref{eq:explicit}. On the physical sheet: 
(i) there is a branch cut starting at $p^2 \geq (m_1+m_2)^2$ (normal threshold) 
(ii) there is no branch cut at  $p^2 = (m_1-m_2)^2$ (pseudo threshold) 
and (iii) there is no singularity for $ p^2 \to 0$ (second type singularity).  
To see the correctness of (ii) one has to note that $\sqrt{\la}$ is imaginary 
for $(m_1 -m_2)^2 < p^2 < (m_1+m_2)^2$. 
In summary it is confirmed  from the explicit representation \eqref{eq:explicit} 
that the pseudo threshold and the second type singularity are, indeed, not present 
on the physical sheet.


\subsection{Cutkosky rules}
\label{sec:CR}

The question of how to compute the actual singularities for physical configurations is answered by 
the cutting rules stated by Cutkosky  \cite{CR} shortly after the Landau equations  were formulated. 
This is by no means accidental as they are closely related. The Landau equations  tell us that 
there is a  singularity if either $q_i^2 = m_i^2$ or $\al_i = 0$ \eqref{eq:LE1} and the Cutkosky rules 
state that the corresponding singularity can be computed by replacing each on-shell (or cut propagator) 
\begin{equation}
\label{eq:CR1}
\frac{1}{q_i^2 - m_i^2 - i0 } \to - 2 \pi i \de^{(+)}(  q_i^2 - m_i^2) \;,
\end{equation}
with the $ \de^{(+)}(p^2-m^2) \equiv  \delta(p^2-m^2)\theta(p_0)$-distribution. 
Before we motivate this rather elegant and surprisingly simple prescription let us state 
the result more explicitly. 
\begin{tcolorbox}
The discontinuity of $I$ \eqref{eq:I}, for real momenta, with propagators $i = 1  \dots r \leq N$ cut is given 
by 
\begin{equation}
\label{eq:CR}
{\rm disc} [I] = (- 2 \pi i )^r \int Dk \frac{ \prod_{i=1}^r \de^{(+)}(q_i^2 - m_i^2)}{\prod_{j=1}^{N-r} (q_{r+j}^2 - m_{r+j}^2)} 
\;,
\end{equation}
and is known as the Cutkosky or cutting rule!
\end{tcolorbox}
Before trying to make plausible the formula  \eqref{eq:CR} let us state the obvious. 
The rule \eqref{eq:CR1} certainly gives the discontinuity of the propagator. The somewhat surprising fact 
is that this seems to be the recipe in any diagram. In the book of Peskin and Schr\"oder \cite{PS}
one can find the bubble graph evaluated in this way.

\begin{figure}
\begin{center}
\includegraphics[width=5.6in]{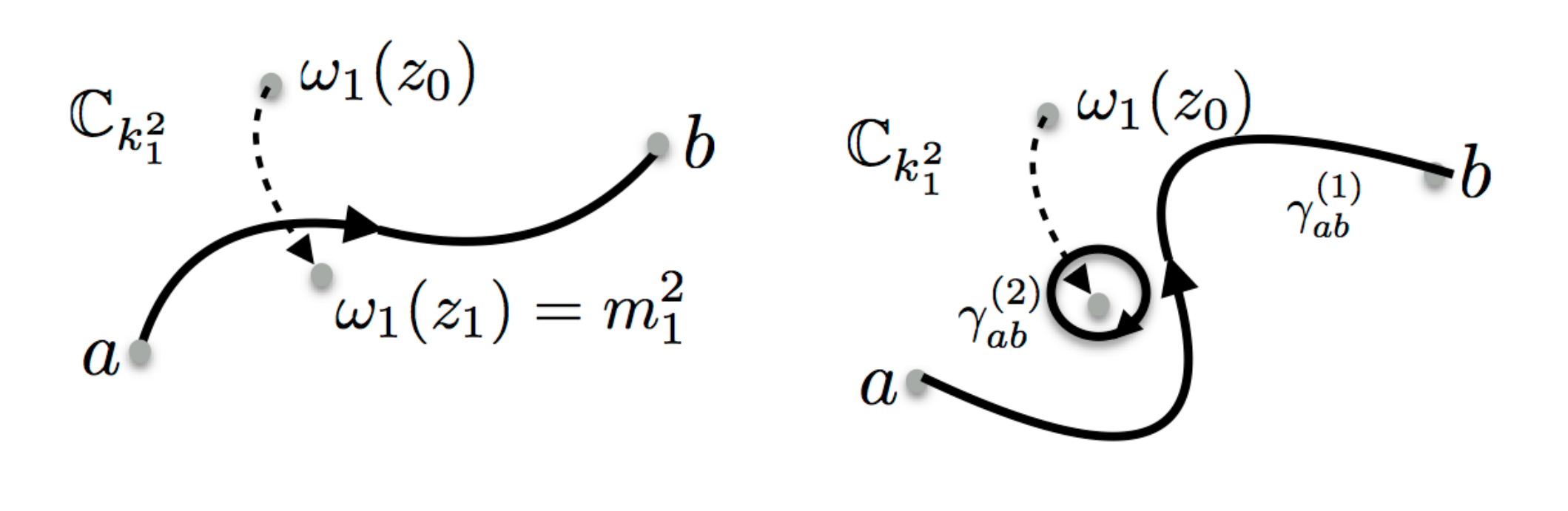} \; 
\end{center}
\caption{\small (left) pinch singularity (right) equivalent path $\ga_{ab} \sim \ga_{ab}^1 + \ga_{ab}^2$ with isolated pinch singularity}
\label{fig:CR}
\end{figure}

In order to motivate the Cutkosky rules we are going to sketch an argument given in the original paper 
\cite{CR} which is also reproduced in \cite{S-matrix}. One considers an integral representation of the form
\begin{equation}
I(z) = \int dk_1^2 \frac{ F(k_1^2,z)}{k_1^2 -m_1^2 - i 0} \;,
\end{equation}
where the variable $z$ is a function of the other momenta external and internal. 
Let the integrand $F$ contain a pole $w_1(z)$ which approaches $m_1^2$ for some $z$ such that there 
is going to be a pinch singularity as shown in Fig.\ref{fig:CR}(left). One then switches to the equivalent configuration 
where the contour is deformed below the mass $m_1^2$ at the cost of encircling the singularity $m_1^2$. In the next step the  integral is performed  using Cauchy's theorem which is equivalent to replacing the 
denominator by           $\de(k_1^2-m_1^2)$.  This argument falls short in justifying the additional physical condition
 $\theta((k_1)_0)$.  Repeated use of the argument above, for each propagator gives the celebrated  Cutkosky rules. 
 The Cutkosky rules have recently been proven more rigorously \cite{Kreimer} using 
 ideas and methods by Pham.
 For the case of normal thresholds (unitarity cuts only)  the method of the largest 
time equation \cite{tHV} provides an elegant derivation of the cutting rules.

\subsection{Anomalous Thresholds \& physical Interpretation of Landau Equations}
\label{sec:fun}

\subsubsection{Brief remarks on anomalous Thresholds} 
\label{eq:termin}

By studying the $1$-loop bubble graph, in  section \ref{sec:normal},  we have encountered  
singularities of the 
normal \& pseudo and 2nd type; 
cf. Tab.~\ref{tab:singularity}.  The most important singularities 
for practical purposes (e.g. dispersion relation, experiments) are the singularities which 
are on the physical sheet of which  the normal one is the only type so far.
This poses the questions whether there are any singularities on the physical sheet.
 By the work of K\"all\'en-Lehmann \eqref{eq:KL} we know that this is not the case  for 
 $2$-point function and we must therefore look at higher point functions.

\begin{figure}
\begin{center}
\includegraphics[width=5.6in]{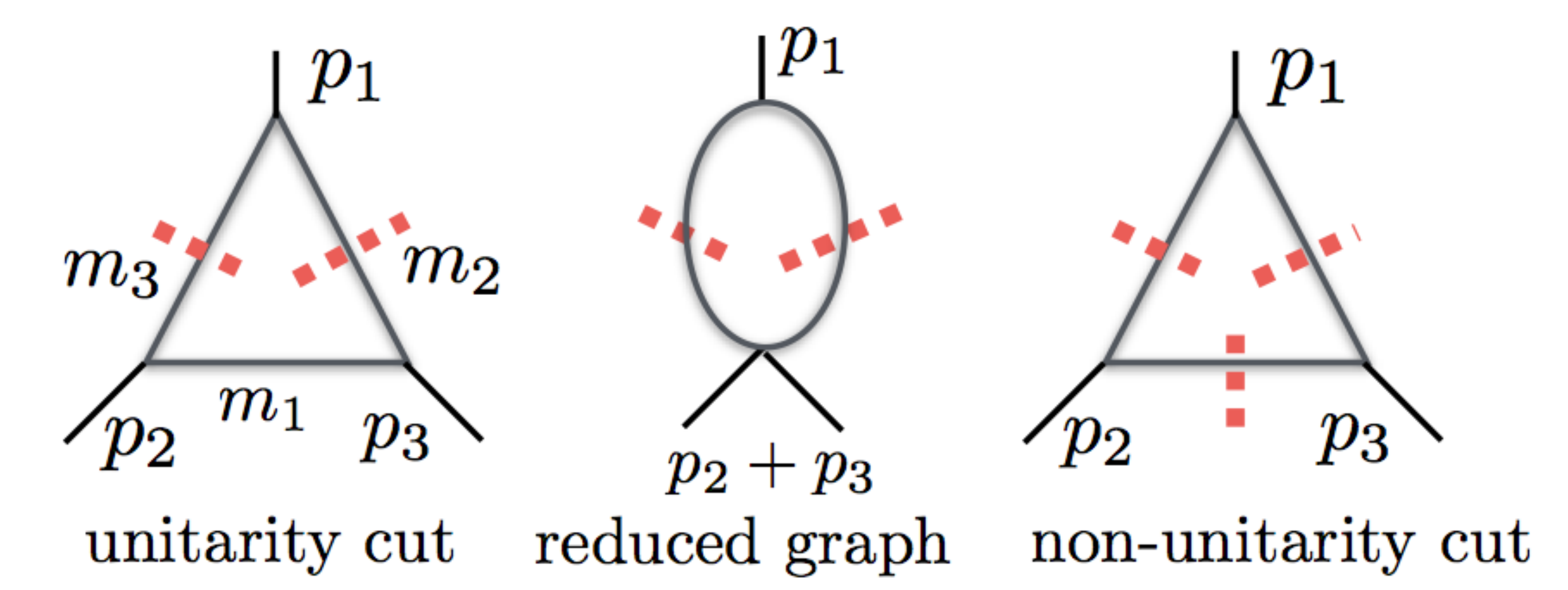} \; 
\end{center}
\caption{\small (left) normal threshold cut (non-leading singularity) 
(middle) the corresponding reduced graph for which $\al_1 = 0$ in the Landau equations. 
 From the reduced graphs all lower singularities can be found. 
(right) the  leading Landau singularity corresponding to a so called anomalous threshold (non-normal threshold)}
\label{fig:three-cuts}
\end{figure}

There are indeed  new classes of singularities for $3$- and higher point functions. 
A $3$-point function Fig.~\ref{fig:three-cuts}(right), for example, can be cut into more than two pieces which is going beyond the singularities discussed so far.
Putting all three   propagators on shell  corresponds to 
$\al_i \neq 0$ at the level of the Landau equations. The corresponding singularities  are known as  \emph{anomalous thresholds}.\footnote{Their existence can be deduced from hermitian analyticity \cite{S-matrix} which in our case corresponds to the property 
that the imaginary part is proportional to the discontinuity.}
It is obvious that the singularities in say $p_1^2$ depend on the values of $p_2^2 $ and $p_3^2$ (provided the line between the two is not 
contracted $\al_1 \neq 0$ as otherwise one encounters the singularities discussed so far).

Whether or not anomalous thresholds appear on the physical sheet, 
returning to our original  questions,  depends on the external momenta 
$p_{1,2}^2$.  
Below we mention examples where the anomalous thresholds appear on the 
physical sheet.  

\begin{itemize}
\item[(a)] Consider the $1$-loop version of the $3$-point function 
with $p^2 \equiv p_2^2 =p_3^2$, $m \equiv m_2 =m_3$ with masses as indicated in 
 Fig.~\ref{fig:three-cuts}(left). For this configuration 
there is an anomalous threshold in $p_1^2 \geq  X $ with $X=   4 m^2 - (p^2 - (m^2 +m_1^2))^2/m_1^2$ 
provided the condition $p^2 > m^2 +m_1^2$ holds  \cite{IZ}.  $X$ is a branch point and 
the higher values are branch cuts.
This anomalous threshold  is below the two particle threshold at $4 m^2$ and 
might be regarded as the very reason for calling these thresholds anomalous!\footnote{This is 
of relevance for form factors which, as we have seen, can be related to $3$-point functions. The anomalous threshold does appear for  the electromagnetic form factors
of the hyperons whereas for for the pion and kaons they do not since global quantum numbers do not allow the condition to be satisfied \cite{IZ}. Appearing and not appearing stands for being on the physical sheet or not.}

\item[(b)] For an example of a momentum configuration where the anomalous threshold is complex and on the physical sheet we refer the reader to the appendix in \cite{DLZ}. This anomalous 
threshold has to be taken into account in the  dispersion relations by choosing  the contour accordingly. 

 The corresponding $3$-point function serves as an example where  
the Schwartz reflection principle \eqref{eq:Schwartz} 
does not apply since the amplitude is imaginary on the 
entire real axis. In some more detail:
 the $1$-loop $3$-point function evaluated in PT
does obey  Schwartz's reflection principle \eqref{eq:Schwartz} with no anomalous threshold on the first sheet. It is though not the correct analytic continuation into the lower half-plane.
Crucially, after elimination of an unphysical branch cut on the real line Schwartz's reflection principle is not obeyed anymore allowing for the anomalous threshold to appear on the physical sheet in the  lower half plane.
\end{itemize}

\begin{table}[h]
$$
\begin{array}{l  ||  l | l | l | l  }
\text{singularity} & \text{normal} & \text{pseudo} & \text{2nd type}  & \text{anomalous} \\ \hline \hline
\text{physical sheet} & \text{yes} & \text{no} & \text{no} &  \text{$p_{2,3}^2$-dependent} \\
\text{remark}  & \text{unitarity} & \text{ }  & \text{mass-indep.} & \text{Leading Landau} 
\end{array}
$$
\caption{\small Different type of singularities described in the text: normal-, pseudo- \& 
anomalous thresholds and 2nd type singularities.
Whether or not the anomalous threshold is on the physical sheet depends 
on the momentum configuration of the other channels. E.g. in the example the triangle graph (cf. Fig.~\ref{fig:three-cuts}(right)) in the text the  $p_1^2$-channels singularities are $p_{2,3}^2$-dependent.
In addition to the singularities mentioned above  there are also, the previously mentioned, mixed type singularities; mixed between 2nd type 
and others for different loop momenta. Not much is known about the Riemann sheet properties of mixed type singularities \cite{S-matrix}.}
\label{tab:singularity}
\end{table}

We end this section with a miscellaneous   remarks on terminology and practicalities.
\begin{itemize}
\item Anomalous thresholds go beyond the concept of unitarity cuts in that they 
allow for cutting the diagram into more than two pieces. Cutkosky \cite{CR} was well aware of this and 
introduced the term \emph{generalised unitarity} in the context of his equations.
\item
The number of propagators that are put on-shell (in a loop) give rise to a natural classification. 
The singularity with the maximal number of on-shell propagators (all $\al_i \neq 0 \Rightarrow  \det Q =0$) is usually referred to as the 
\emph{leading Landau singularity}. In the case of the triangle diagram the anomalous threshold is the leading Landau singularity.
\item
When all but one external momenta are kept below the thresholds  there are only normal thresholds  on the physical  sheet in the corresponding four momentum squared. 
For analytic computations
this constitutes a method for obtaining correlation functions in a certain kinematic range. 
The remaining domain is then obtained by analytic
 continuation.\footnote{This makes use of the analyticity postulate of the S-matrix theory 
 in the that 
the domain of analyticity is the maximal one consistent with 
unitarity (normal thresholds) and crossing symmetry. Crossing symmetry means that
if scattering $A + B \to C +D$ and the decay $A \to \bar{B} +C +D$ 
are both possible then they are described by the same amplitude through analytic continuation. 
These postulates are seen to be correct in concrete QFT computations and 
believed to be true in general.  Some results are known for 
 $3$- and $4$-point function through the work of K\"all\'en \cite{three,four}  
and even more generally from tools like the Edge of the Wedge theorem \cite{IZ,PCT}.}

For the assessment of dispersion relation, outside the range of concrete computations,  
 this is not a practical method since one needs to know the location of all singularities on the physical sheet in order to choose a path $\gamma$ which does not cross 
any singularity. 
\end{itemize}

\subsubsection{Physical Interpretation of the second Landau Equations (\ref{eq:LE2},\ref{eq:LE2b})}

For physical momenta  Coleman and Norton have given an interpretation 
of the second Landau equation \cite{CN}.  It is found  that (\ref{eq:LE2},\ref{eq:LE2b}) ensures  that the corresponding diagram can occur as a real process where the Feynman-parameter $\al_i \sim \tau_i/m_i$ 
has the interpretation of being  the proper time $\tau_i$ divided by the mass of the 
propagating particle. 

This means that  $\al_i =0$   corresponds to the case where the $i^{\rm th}$ particle
does not propagate at all and gives the reduced graphs (e.g. Fig.~\ref{fig:three-cuts}(middle)) a more direct meaning.
At last we note that the Coleman-Norton interpretation is a very  reassuring result 
in view of the optical theorem's \eqref{eq:optical}  
statemant that the discontinuity (related to singularities), of the forward scattering amplitude, originates from  physical intermediate states.

\section{Outlook}

Even though dispersion relations are an old subject and a pure dispersive approach to particle 
physics has proven to be too complicated in practice, 
dispersion theory is and will remain a powerful tool in QFT as it follows 
from first principles and is intrinsically non-perturbative.  This makes it particularly useful 
for hadronic physics which is not directly accessible by a perturbation expansion in the strong coupling constant. Dispersion relations are the most solid approach to quark hadron duality. Any approach to quark hadron duality should either start from or connect to dispersion relations. 

In recent years dispersion relations and unitarity methods 
have also seen a major revival in evaluating 
perturbative diagrams  (e.g. \cite{Dixon,SAbook,Henn,RT} for  reviews and applications). 
The bootstrapping programme has witnessed new 
exciting developments  by limiting/fixing the target space functions of amplitudes. 
Old tools such as the Steinman relations, which are physical conditions 
on double discontinuities, have led to promising simplifications valid beyond perturbation 
theory \cite{boot}.

Furthermore dispersion relations can serve 
to prove positivity, for example, when a physical quantity can be expressed as an unsubtracted
dispersion integral with positive integrand (discontinuity).
Examples are   the so-called $c$- and $a$-theorems, which characterise the  irreversibility 
of the renormalisation group flows in 2D and 4D. The dispersive proofs are given in 
\cite{Cappelli:1990yc,KS11} in two and four dimensions by looking at two and four-point functions respectively. On another note, positivity of the K\"all\'en-Lehmann representation seemed to exclude the possibility of asymptotically free gauge theories in  1970 
\cite{Wilson3}. The Faddeev-Popov ghosts (negative metric) proved to be the loophole 
in this argument as they give rise to the negative sign of the $\be$-function   \cite{AF1,AF2}. 

I am grateful to James Gratrex for proofreading and to Einan Gardi for discussions on second type singularities. Apologies for all  relevant references that were omitted.  
Last but not least I would like to thank the organisers of the ``Strong Fields and  Heavy Quarks" as well 
as the participants for a stimulating and pleasant atmosphere. I really did  enjoy my trip to and around Dubna! 
  
 \appendix

\section{The Schwartz Reflection Principle}
\label{app:Schwartz}

Consider an analytic function $f(z)$ with $f(z) \in \mathbb{R}$ for $z \in I_R$ where $I_R$ is an interval 
on the real line. Then the following relation holds
\begin{equation}
\label{eq:2c} 
f(z) = f(z^*)^*  \;,
\end{equation}
which can be analytically continued to the entire plane. Note that analytic continuation is unique from 
any set with an accumulation point for which an interval is a special case.  Hence Eq.~\eqref{eq:2c} implies that
\begin{equation}
{\rm Re}[ f(z)] = {\rm Re}[ f(z^*)]   \;, \quad {\rm Im}[ f(z)] = - {\rm Im}[ f(z^*)]  \;.
\end{equation}
Choosing $z = s + i0$ with $s \in \mathbb{R}$ it then follows that 
\begin{equation}
\label{eq:Schwartz}
 {\rm disc}[ \Gamma(s)]  =    2i  {\rm Im}[ \Gamma(s)]  \;,
\end{equation}
which is a result known from experience with 2-point functions and intuitively in  accordance with the optical theorem.

\section{Conventions}
\label{app:conventions}

Here we summarise a few conventions. We are using the Minkowski metric of the form 
\begin{equation}
g_{\mu \nu} = {\rm diag}(1,-1,-1,-1) \;,
\end{equation}
the following abbreviations for integrals over space and momentum space
\begin{equation}
\int_x = \int d^4 x  \;, \quad  \int_k = \int d^4 k \;,
\end{equation}
and the relativistic state normalisation
\begin{equation}
 \vev{p|p'} =2 E_p  (2 \pi)^3  \de^{(3)}(\vec{p} - \vec{p'}) \;,
\end{equation}
where $E_p = \sqrt{\vec{p}^2 + m^2}$ with $p = (E_p ,\vec{p})$.

\bibliographystyle{utphys}
\bibliography{input3}

\end{document}